\documentclass[12pt,letterpaper]{article}
\usepackage{amsmath,amssymb,pgf,pgfarrows,pgfnodes,float,appendix, hyperref}
\usepackage{graphicx}
\usepackage{subfigure}

\usepackage[margin=0.9in]{geometry}
\usepackage{pgfplots}
\usepackage{amsmath}
\usepackage{tikz}

\def\slash#1{\ensuremath{\;/\!\!\!\! #1}}
\title{{\rm\footnotesize \qquad \qquad \qquad \qquad \qquad \ \qquad \qquad \qquad \ \ \ \ \ \                  RUNHETC-2019-15 }\vskip.5in    On the Limits of Effective Quantum Field Theory: Eternal Inflation, Landscapes, and Other Mythical Beasts }
\author{Tom Banks \\
Department of Physics and NHETC\\
Rutgers University, Piscataway, NJ 08854\\
E-mail: \href{mailto:tibanks@ucsc.edu}{tibanks@ucsc.edu}}

\date{}
\begin{document}
\maketitle

\begin{abstract}
We recapitulate multiple arguments that Eternal Inflation and the String Landscape are actually part of the Swampland: ideas in Effective Quantum Field Theory that do not have a counterpart in genuine models of Quantum Gravity.
\end{abstract}

\section{Introduction}

Most of the arguments and results in this paper are old, dating back a decade, and very little of what is written here has not been published previously, or presented in talks.   I was motivated to write this note  after spending two weeks at the Vacuum Energy and Electroweak Scale workshop at KITP in Santa Barbara.  There I found a whole new generation of effective field theorists recycling tired ideas from the 1980s about the use of effective field theory in gravitational contexts.   These were ideas that I once believed in, but since the beginning of the 21st century my work in string theory and the dynamics of black holes, convinced me that they were wrong.   I wrote and lectured about this extensively in the first decade of the century, but apparently those arguments have not been accepted, and effective field theorists have concluded that the main lesson from string theory is that there is a vast landscape of meta-stable states in the theory of quantum gravity, connected by tunneling transitions in the manner envisioned by effective field theorists in the 1980s.  

On the contrary, I've been arguing for almost 20 years that the actual mathematical evidence from string theory shows that the effective field theory notion of meta-stable (and stable) vacuum states of a quantum field theory model is not applicable to models of quantum gravity.  The most rigorous parts of the evidence for this claim come from mathematical definitions of models of quantum gravity in terms of Old Matrix Models, Matrix Theory, and AdS/CFT.   This evidence reinforces arguments that I developed based on the Coleman DeLuccia theory of gravitational tunneling, and black hole physics.

The most general way of stating the main issue is the following.  Consider a low energy effective field theory Lagrangian, including gravity.  If we consider the metric to be a fixed classical background and quantize the other fields, then, if the spatial sections of the metric have finite volume\footnote{The finite volume of spatial sections depends, of course, on the time slicing.   Usually some preferred slicing, in which there's a {\it e.g.} a time independent Hamiltonian formulation of the model, is used to define the spatial sections.} the system has a unique ground state.  Gaussian states centered around all possible field configurations are normalizable states in the Hilbert space.   If we take the volume to infinity, then states settle into different orthogonal sectors of the Hilbert space, such that no finite product of local operators acting on the ground state in one has an overlap with
the ground state in another.  These are characterized by different ground state expectation values for scalar fields, for ground states of equal energy density.  In QFT these states are identical in the UV, and differ only in their low energy spectra.

There are also meta-stable states associated with minima of the scalar field potential 
with non-minimal energy density.  Transitions between such states and the true minima occur, within the semi-classical regime, by nucleation of critical bubbles in whose interior the scalar is equal to its true minimum.  Critical bubbles then expand rapidly, with a bubble wall velocity approaching that of light.  The interior of the bubble settles down and behaves just like the true vacuum, over an ever expanding volume.  Every localized object that is an excitation of the false vacuum is eventually engulfed by a bubble.

In fact, multiple bubbles are always nucleated, and they collide and merge.  Every expectation value of products of finite numbers of local operators at fixed finite points in space-time quickly becomes equal, up to small and diminishing corrections, to its value in the true vacuum.

A naive translation of these field theory ideas to hypothetical models of quantum gravity leads to the idea of eternal inflation (EI)\footnote{There are two versions of EI, "slow roll" and "tunneling". The criticisms based on field theory in dS space apply to both, those based on Coleman De Lucia instantons only to the tunneling form of EI.
T. Rudelius\cite{rudelius} has recently given a critique of slow roll EI.}.   In a classical Lagrangian with dynamical gravity, a stationary point of an effective potential, with positive energy density, gives rise to a de Sitter solution.  If the solution is not at the lowest point on the potential then we expect this solution to be meta-stable.  It should decay by bubble nucleation.  Classical dS space in global coordinates is, after the mid-point of global time, a superluminally expanding 3-sphere.   EI is based on the claim that there is an equal probability per unit 4 volume for a tunneling event to occur at any point on this manifold.  We will argue that neither quantum field theory in de Sitter space, nor the semi-classical theory of quantum gravitational tunneling due to Coleman and De Luccia, lends any support to this assumption.

The other key idea, also dating to the 1980s, that we will examine is the proposition of a Landscape of Vacua, corresponding to different minima of a low energy effective potential.   This idea was incorporated into string theory, originally in an attempt to show that different string theoretic models were just different states in a single quantum theory, with the continuous parameters that characterize those models corresponding to the expectation values of fields in a low energy Lagrangian.  While the latter phrase is literally true, in a model of quantum gravity it does not imply that the different solutions correspond to different quantum states with the same Hamiltonian.  There are indeed moduli spaces of supersymmetric string models, but different points in the moduli space do not share the same high energy behavior.  In addition, one cannot create a state with an arbitrarily large region with one value of the moduli by scattering  experiments done at another value, unless the two points in moduli space are close to each other.  Field excursions of more than the Planck scale form a black hole around the bubble of "other vacuum".

Furthermore, even if two points are close to each other in moduli space, high energy scattering amplitudes at a range of impact parameter that grows with the energy are dominated by production of large black holes, whose decay amplitudes are sensitive to details of the low energy spectrum, and therefore differ in the two models.
Different points in moduli spaces of string models are NOT different states of an underlying quantum Hamiltonian.  Neither the high nor low energy spectra of the Hamiltonians are close to each other.

A refinement of this observation is that different solutions of a given low energy Lagrangian correspond to different states in the same quantum theory of gravity only when their large distance space time geometry is identical.  The reason for this is that the high energy spectrum of any model of quantum gravity is dominated by black holes, and the spectrum of black holes depends on the asymptotic spacetime geometry.

We will give numerous examples of this phenomenon, the most prominent of them coming from the AdS/CFT correspondence, where it has been evident from the very beginning, despite the community's decision to ignore it.  To give just two of many examples
\begin{itemize}

\item The Lagrangian of Type IIB SUGRA in ten dimensions has a two parameter family of $AdS_5 \times S^5$ solutions.  The two parameters are the cosmological constant and the value of the complex dilaton field at infinity. AdS/CFT shows us that the c.c. in Planck units is actually quantized. Each quantized value corresponds to a {\it different} conformal field theory, with different spectrum of high dimension operators.  The second parameter is complex and corresponds to a {\it conformal manifold} of CFT models, with different spectra.  For large values of the complex parameter, the model becomes weakly coupled ${\cal N} = 4$ $ SU(N)$ Super Yang Mills theory.  The c.c. in Planck units is a function of $N$ and the "dilaton expectation value" is the complex 't Hooft coupling parameter $\lambda^{-1} = \tau / N$.  The spectrum of operators depends on both $N$ and $\lambda$ for both low and high dimension operators.   It's only at large $\lambda$ and $N$ that a gravitational description is appropriate.
There's also a Minkowski solution, which is not a CFT.

\item Consider any Lagrangian for gravity coupled to matter including scalar fields, with a stable negative c.c. stationary point of the scalar potential.  Such models are candidates to be dual to CFTs.  Any such model, has an infinite set of other solutions, which are FRW universes with a time dependent scalar initially starting at some non-stationary point of the potential.   A generic solution of this type has a negative c.c. Big Crunch\footnote{This was first pointed out by Coleman and DeLucia\cite{cdl}. The FRW coordinates for $AdS$ space have coordinate singularities.  For generic initial conditions the equations for scalar fields with any potential, preserving the FRW symmetries, become truly singular at those points and lead to the Big Crunch gravitational back reaction.}, and has no connection to the AdS solution or to a CFT.  Solutions of the effective field equations with a prescribed behavior at Euclidean AdS infinity (equivalently, spatial infinity in global coordinates) are interpreted either as states in the quantum theory, or perturbations of the CFT by some local operator.  The FRW Crunch solutions obey neither of these boundary conditions and have nothing to do with the CFT which describes a complete theory of quantum gravity in AdS space.

\end{itemize}

\section{Instantons in the Presence of Curved Geometry and Dynamical Gravity}

The semi-classical calculational tool that is used in the analysis of the decay of meta-stable states is the method of instantons, or imaginary time solutions of the classical equations of motion.  In field theory in Minkowski space, an instanton is a smooth finite action field configuration whose asymptotic scalar fields approach their value at some local minimum, which is not the absolute minimum.  One can prove that the minimal action configuration is always rotationally symmetric.  The radial derivative of the instanton must vanish at the origin, in order for the configuration to be smooth and satisfy the boundary condition at infinity.  For definiteness, let us imagine a single scalar field in $d$ space-time dimensions, with a potential having two local minima.  The imaginary time field equations are
\begin{equation} \phi_{rr} + \frac{d - 1}{r} \phi_r  - V^{\prime} (\phi ) = 0 . \end{equation} Upon analytic continuation to Minkowski space, the point $r = 0$ becomes the boundary of a light cone.  If we take it to be future pointing, the analytic continuation of the solution is constant on space-like hyperbolae  and describes the future evolution of the expanding critical bubble.  

The effect of curved space-time geometry becomes apparent even if we do not treat the geometry as a dynamical variable.  In order to do the analytic continuation, we need to consider a space-time with time-like Killing vectors.  An example is de Sitter space, the surface of a timelike hyperboloid in $d + 1$ dimensional Minkowski space.  When we make time imaginary, the hyperboloid becomes a sphere.  de Sitter space is the maximally symmetric space-time with positive cosmological constant, so in a theory with dynamical gravity it would seem to be the right venue for discussing the decay of a false vacuum state with positive energy density.  

We first want to discuss the quantized scalar field in dS space, which is a perfectly well defined problem.  Nonetheless, since adding a constant to the potential does nothing when gravity is non-dynamical, we will take the dS cosmological constant to be equal to the value of the potential at the false minimum.  The maximally symmetric instanton is a function only of the polar coordinate on the $d$ - sphere and is invariant under $SO(d)$.
When the true and false minima are close in energy density, the thin wall approximation\footnote{The thin wall approximation has a scalar field configuration that is a smoothed out approximation to $\phi_t \theta (a - r) + \phi_f \theta (r - a)$, where $\phi_{t,f}$ are the two minima of the potential.} is valid, and if the potential varies slowly on the Planck scale, then the gravitational corrections to the equation for the instanton are small.  There are however two new features of the solution that cannot be neglected.  First, we note that the gradient of the scalar is a vector field on the sphere and it must vanish at two different points in polar angle.   At each of these points we can analytically continue the solution to Euclidean time and there are two real time classical solutions, which evolve to the two local minima of the potential\footnote{Actually, if the lower minimum is at negative energy the gravitational corrections are not negligible, as mentioned above.  The solution evolves to a Big Crunch, near which the scalar field varies over its full range of values.}.  The second peculiar feature of the dS problem is that all of the instanton collective coordinates are compact.  We'll discuss the implications of both of these facts below.

The analytical continuation of dS space to imaginary time is the $4$ sphere.  If we choose to analytically continue an isometry of the sphere we get a static coordinate patch of dS space, the maximal causal diamond along any time-like trajectory in the space-time.  The periodicity of the isometry direction implies that the analytically continued Euclidean correlation functions are thermal, with the Gibbons-Hawking\cite{gh} temperature.  There is another copy of this causal diamond, which can be obtained by performing the antipodal map on the sphere before continuation.  In the Penrose diagram of Figure 1, these are regions $I$ and $II$.  Every event in the global dS geometry is causally related to events in either of these two diamonds, but as shown in the figure, the time in the two diamonds runs in opposite directions.   A similar situation for the Kruskal extension of the black hole solution in Minkowski space motivated W. Israel\cite{israel} to claim that that spacetime was a representation of the "thermofield double" (TFD) of the quantum system corresponding to the one sided black hole.
\begin{figure}
\begin{center}
\includegraphics[scale=0.5]{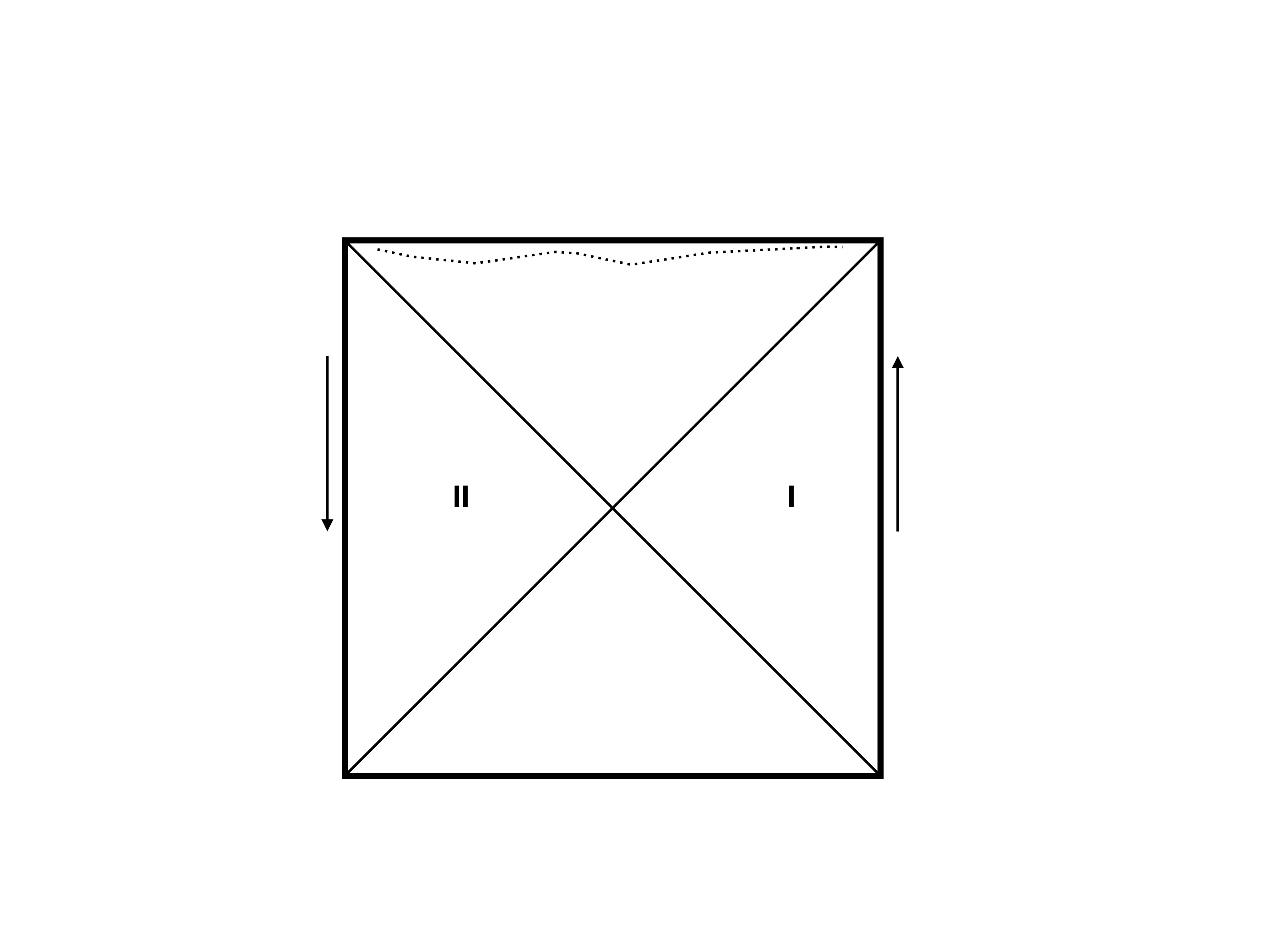}
\caption{The Penrose Diagram of de Sitter Space showing the two copies of a maximal causal diamond, whose opposite global time orientation leads to the thermofield double interpretation of the spacetime.  Also shown is a line of late time insertions in the region exterior to the two diamonds.  These are causally connected to the future boundary of region I and the past boundary of region II.  If the late time slice is pushed to infinite global time, with a constant global density of insertions, then this corresponds to an infinite number of insertions on the cosmological horizon and creates a singular state orthogonal to every state in the Bunch-Davies Hilbert space.
}
\label{l}
\end{center}
\end{figure}

The TFD is a purification\footnote{Any density matrix on a Hilbert space ${\cal H}$ can be thought of as coming from tracing over the degrees of freedom in Hilbert space ${\cal H}_1$, of an entangled state in the tensor product Hilbert space ${\cal H}\otimes {\cal H}$. Any such representation is called a purification of the density matrix.}
 of the thermal density matrix by entangling it with a time reversed copy of the same system.  Maldacena\cite{malda} generalized Israel's conjecture to black holes in AdS space, where one can subject it to many tests.  It passes them with flying colors.  This motivated the authors of \cite{susskindetal} to propose the same interpretation for dS space.  That is, global dS space should be thought of as the thermofield double of a maximal causal diamond.   These are however, strictly speaking conjectures about a hypothetical quantum theory of gravity.  For the next few paragraphs we will simply discuss quantum field theory in dS space.  

We can define this system in terms of the scalar field functional integral on the sphere.  The models we are studying have no zero mode divergences because the curvature of the scalar potential is positive at both of its minima.  Analytic continuation of the functional integral defines thermal correlators in the causal patch.  Transformations of the de Sitter group $SO(1,4)$ enable us to continue those correlators so that the points lie anywhere in dS space\cite{nappietal}.   The spatial volume of the causal diamond is finite, so that any questions about points at infinite spatial separation in global coordinates can be answered by appealing to group theory and the physics of a finite system.   For free field theory one can verify that these dS continued correlators are expectation values in a pure state of a canonical quantum field theory defined on the global space-time, with some choice of a global dS time coordinate (the choices are related by dS transformations).  This is called the Bunch-Davies vacuum, and we will retain that name for the generalization of these computations to interacting theories.

We can obtain a potential with two unequal minima using cubic and quartic interactions.  In two and three dimensions the resulting field theory has a rigorous mathematical definition\cite{nappietal}, while in four dimensions the perturbative theory needs only renormalizations of the mass, coupling and $R \phi^2$ coefficient. Since $R$ is a constant and we are not considering variations in the dS background, the latter term is part of the mass renormalization.  The theory is not asympotically free in four dimensions, but this introduces only issues logarithmic in the UV cutoff  and does not really affect our analysis of instanton dynamics.

The question now is what the instantons in this model mean.  Brown and Weinberg\cite{bw} have presented very convincing arguments that they represent the expected transitions between meta-stable states of the finite volume system.  Transitions back and forth between meta-stable states at finite volume are expected at finite temperature.  We can now ask what the significance of the finite number of instantons allowed on the sphere is, from this point of view.  The answer is that, within a causal diamond, the only instantons that can effect correlation functions in the interior of the diamond, are those that nucleate near enough to the geodesic through the diamond that their effects can be felt before the nucleated bubbles are swept "out of the horizon". From the point of view of the geodesic time coordinate in the diamond, nothing goes out of the diamond, so the phrase in quotes should really be replaced by "becomes indistinguishable from a random thermal fluctuation on the horizon".
In this way of interpreting the quantum mechanics of field theory in dS space, instantons outside the causal diamond are just dS transforms of those that happened to occur within the diamond and they have no effect on physics within the diamond.  This means that in the Bunch-Davies state the data on the frequency of transitions between the two metastable states collected by a detector traveling along any timelike trajectory will be identical for two trajectories related by an $SO(1,4)$ transformation.  The predictions for geodesics are simply thermal statistics, while for other trajectories they will be distorted by the motion relative to the thermal bath.  They are completely unambiguous and suffer only from the conventional UV problems of $\phi^4$ theory in the $4$ dimensional case.  These can be removed by choosing the scalar to be a Brout-Englert-Higgs field of an appropriate asymptotically free gauge theory.  

Adherents to the philosophy of Eternal Inflation (EI), should be disturbed at this point.  Our model is the basic building block of EI, but a standard field theory analysis of it leads to none of the peculiar predictions nor the ambiguities of EI.  EI theorists assume that the "exit points" of tunneling events can occur at arbitrary points in the expanding phase of the dS manifold, as viewed along some particular set of global timeslices, which cut both of the slicings of the thermofield double interpretation, with the {\it same} time orientation for both of them.  The instantons at different points along the slice are assumed to occur with statistically uncorrelated probabilities.   This is not the case for the Bunch-Davies vacuum in which the instanton distributions outside the causal diamond of one observer are just unitary transformations of those inside the diamond.

One can get a flavor of what the EI {\it ansatz} is probing, by inserting operators of the form 
\begin{equation} \prod_P e^{- (\phi (f_P) - \phi_{\pm} (P))^2 /2 \Delta^2 } . \end{equation} into the B-D vacuum.  Here $P$ labels points along some global time slice, separated by a bit more than a Hubble length, and $f_P$ is a smooth function of compact support in a region of spacetime surrounding $P$ and having space and time extents of order the Hubble length.  $\phi_{pm} (P)$ are the two minima, multiplied by the volume of the spacetime region around $P$.  These operators test whether the region around $P$ resembles 
one or the other of the vacua, if $\Delta$ is chosen much smaller than the field space distance between the minima.  EI theorists are studying the insertions of such operators along each expanding slice in such a global slicing in the limit that the global time goes to infinity, for all possible choices of which minimum is tested at each point $P$.   It is little wonder that the results are ambiguous and slice dependent.  Furthermore, we can use the Heisenberg equations of motion to relate these expectation values to ones where all of the test operators are inserted on either the past boundary of one thermofield double diamond, or the future boundary of its "twin", for any choice of the static coordinate system.   The infinite set of insertions demanded by EI makes these two boundaries singular, again making it clear that EI is not talking about the Hilbert space based on the BD state.  States in which we've made the "measurements" to test the predictions of EI are orthogonal to any state made by acting with finite numbers of fields on the BD state, while such finite product states are dense in the Hilbert space.

A final way of understanding what EI is doing incorrectly uses the observation that the collective coordinates of the instanton gas form a compact space.  EI computations instead use the full $SO(1,4)$ group as collective coordinates.  This mistake is analogous to that made in some of the earliest instanton computations for false vacuum decay in Minkowski space\cite{russians}, which used the Lorentz group rather than the Euclidean group as the space of collective coordinates.  This error was corrected in Coleman's seminal paper\cite{fate}.  

In summary, in the quintessential false vacuum EI model, treated as a scalar quantum field theory in a fixed dS space, computations of correlation functions of finite numbers of fields in the BD vacuum show no signs of the infinities or ambiguities of EI computations.  The predictions for the number of expected transitions between true and false vacua along any timelike geodesic are unambiguous and thermal.  For any other timelike trajectory they are distorted in a finite way by the trajectory's motion w.r.t. the rest frame of the thermal bath.  
Theorists who want to stick to their EI guns, have to imagine that somehow an as yet unknown quantum theory of gravity changes these field theoretic results.  We will now demonstrate that, in fact, effective field theories of gravity make it appear even less likely that the speculations of EI are correct.

\subsection{Gravitational instantons for transitions between dS minima}.

The technical aspects of instanton solutions to the coupled Einstein-scalar equations are well covered in many places\cite{gravinstrefs} so we will just summarize the results.  The Riemannian geometry of the instanton describing transitions out of a dS state is an ovoid whose cross section, in four dimensions is a $3$ sphere.  Technically, this is a three sphere bundle over an interval, with three spheres of vanishing radius at the ends of the interval. If $\theta \in [0,\pi] $ is the coordinate in the interval, then the gradient of $\phi$ must vanish at the endpoints.  These are the points of analytic continuation to Lorentzian time, and the continued solutions are negative curvature FRW metrics, which locally asymptote to one of the two dS minima.  For potentials that are sufficiently slowly varying functions of $\phi/ m_P$, the solution does not exist.  $\phi$ must sit at the maximum between the two minima and the three sphere has the radius of the dS space with that value of the c.c. .  This is called the Hawking-Moss instanton and it always exists, even when the potential supports ovoidal instantons.  One must check whether the Hawking-Moss instanton has higher or lower action than the ovoids to compute the dominant transition rate.  It's possible that models with no ovoid solutions are part of the Swampland\cite{vafa} but we will not use this speculation.

A new feature of gravitational instantons with a dS false minimum is that the gravitational contribution to the action is negative and always dominates.  The instanton action is negative.  To resolve this we must subtract the negative action of the pure dS solution, either of the true or false vacuum.  This always gives a positive number and so defines two probabilities $e^{- (S_I - S_{1,2})}$, which are interpreted, as in the Brown Weinberg paper, as the transition rates between the two minima.  Note that in the limit studied in the previous subsection, where the scalar potential is much smaller than Planck scale and varies on field scales much smaller than Planck, $S_1 \sim S_2 \sim S_I$.  The probability for transition from the false to true vacuum is approximately given by the exponential of a field theoretic instanton, while the reverse transition is Boltzmann suppressed unless the difference in vacuum energy densities is of order the dS temperature or smaller. 

In\cite{heretics} I pointed out that one could interpret the ratio of transition rates $e^{ - (S_1 - S_2)} $ as an expression of the Principle of Detailed Balance, because of the fact that the Euclidean action of dS space is equal to the Gibbons-Hawking entropy of that space.  The principle of detailed balance relates a ratio of transition probabilities to a difference in free energies rather than entropies.  However, if the expectation value of energy is of order the temperature, and the entropy is large, then the difference in free energies is the difference in entropies, up to a factor $(1 + c/\Delta S)$ where $\Delta S$ is the entropy difference.  One is thus led to the conclusion that dS space transitions resemble those in a finite entropy system, where nothing ever decays irreversibly.  The false vacuum is a much lower entropy sub-system of the true vacuum Hilbert space, which is nonetheless visited on classical recurrence time scales.  

Thus, these semi-classical clues to the correct theory of quantum gravity support and reinforce the conclusions we drew from rigorously defined quantum field theory.  The infinite number of independent operators hypothesized in the theory of EI, simply cannot be relevant to the physics of a finite entropy system.  One may ask how those operators appear in QFT quantized on global time slices of dS space.   If we recall that those time slices are causally connected to the two finite sized causal diamonds of antipodal timelike geodesics, it is easy to see that the independent long wavelength modes on a late time global time slice, are related by Heisenberg evolution to modes of arbitrarily short wavelength in the diamonds.   This criticism of field theoretic treatments of dS space has been made many times, and some authors have claimed that a proper treatment will lead to modifications of the inflationary predictions for CMB fluctuations\cite{UVclaims}.  Those hypotheses were debunked by appealing to models with coordinate dependent UV cutoffs, which kept all modes of the field theory, but modified their action\cite{debunkUVclaims}.   It has been known (or should have been) since the work of Cohen, Kaplan and Nelson\cite{ckn} that no conventional UV cutoff of field theory captures the obvious physical requirement that one must be skeptical of states in QFT whose gravitational backreaction would create a black hole of size larger than the support of the operators that create the state from the QFT vacuum.  As we'll see in more detail below, a mathematically precise formulation of this criterion\cite{bousso} is that the entropy of the quantum system in a causal diamond is bounded by the area $A$ of the {\it holographic screen} of the diamond.  A covariant form of the CKN bound is that the field theory entropy should be smaller than \begin{equation} c (A)^{\frac{d - 1}{d}}\end{equation} where $c$ is a constant of order one. This principle is the basis of a conjectural general theory of quantum gravity called {\it Holographic Space Time}\cite{hst}, but in this note we'll use it only as a rough guide. When combined with unitarity of quantum time evolution, the Covariant Entropy Principle (CEP) rules out the possibility that the infinite set of independent commuting operators hypothesized in EI can evolve from the degrees of freedom in a finite area causal diamond.

\subsection{Gravitational instantons for transitions between Minkowski or dS space and a negative c.c. Big Crunch.}

Let us begin by stating unequivocally that there is no such thing as a transition from Minkowksi of dS space to AdS space.  The seminal paper\cite{cdl} explained why generic solutions for gravitational tunneling "to a negative c.c. region of the potential" are really transitions to a negative c.c. Big Crunch. The scalar field does not remain in the basin of attraction of the negative energy minimum, and in the QFT approximation the entropy of the system appears to increase without bound.  Some authors, consider the Big Crunch to be a "terminal vacuum", to which the initial dS or flat
space decays without possibility of recovery.  We will see that there are two classes of potentials that have gravitational instantons for transitions between a flat or dS space and negative c.c. Big Crunches.   Neither of them has the interpretation of a terminal vacuum.

To orient ourselves, we begin with transitions from flat space to negative c.c. crunches.  One of the most interesting results of the CDL paper was their observation that, within the thin wall approximation, some transitions that would have been decays of a meta-stable vacuum in Minkowski quantum field theory, did not occur if the negative vacuum energy density were small enough.  This occurs because volume scales like area in AdS space, so the existence of critical bubbles of true vacuum depends on parameters in the Lagrangian, and is not guaranteed by geometry.  This quite general phenomenon, is related to the general relativistic positive energy theorem\cite{YauWitten}.   If a gravitational Lagrangian has a locally stable Minkowski solution, then one can often prove a theorem showing that every localized field configuration has positive energy.   In fact, one can show\cite{abj} that given any Lagrangian with a finite number of scalar fields and a potential $V$ with a finite number of negative energy density minima, there is a one parameter deformation $a \delta V$ such that for sufficiently large $a$ one obtains a potential with a positive energy theorem.   Indeed if we take $\delta V$ to be a smooth non-negative function with compact support only in the regions where the original potential is negative, it is obvious that this is true.  The CDL examples show that there can even be a positive energy theorem when the full potential still has negative regions. The work of\cite{abj} exhibited smooth solutions, beyond the thin wall limit, with this property.  Indeed, it's likely that any potential of the form $M^4 V(b \phi_i /m_P)$ will have a positive energy theorem for $b$ less than some $b_0$ of order one.     The infinite dimensional space of all possible potentials with a locally stable zero energy minimum thus has a wall of co-dimension $1$, above which the potentials all have a positive energy theorem.  We call this wall The Great Divide.

Now let's take a potential above the Great Divide and add a small positive constant $v_0$.  There is always an instanton that mediates the transition from dS to the negative c.c. region, at least the Hawking Moss instanton.  In the limit of small $v_0$ the absolute value of the instanton action is much less than the dS action, which is equal to the dS entropy.  The transition rate thus has the size one might expect for a transition from a high entropy to a low entropy state, analogous to a transition to a {\it higher} energy dS space.  Moreover, the maximal causal diamond in the singular Big Crunch space time to which the transition occurs has a small area, so that using the Covariant Entropy Bound\cite{fsb} one concludes that there is much less entropy in this region than in the dS state.   In the case of the instanton for dS to dS transitions  one concluded that the ratio of transition probabilities was the exponential of the difference of the entropies (defined as the area in Planck units, divided by $4$).  We can apply the same principle to the singular space time and conclude that this "decay" is really a rare transition to a low entropy state, which will be followed by a transition back on a much shorter time scale.  This is analogous to the fabled transition in which all the air in a room spontaneously collects in one cubic centimeter.  Note that this is much more consistent with continuity between the positive and zero c.c. results than the interpretation of this transition as a decay.  The decay interpretation requires us to believe that a completely stable state can be completely destabilized by addition of an infinitesimal positive constant to the Lagrangian.  Even more generally, unitarity in quantum mechanics implies that every transition between two quantum states is invertible.  The word {\it decay} always signifies that the ensemble of decaying states has much less entropy than the ensemble of decay products.

The interpretation of the negative c.c. Crunch as a "terminal vacuum" never made much sense.  The contracting phase of the Big Crunch acts as anti-friction in the scalar field equations of motion.   Before the singularity is reached, the scalar's kinetic energy grows larger than the barrier between the two minima and the field does not remain in the basin of attraction of the negative minimum.  If the sectional curvatures on field space are negative (as is universally true on moduli spaces of string theory), the motion is chaotic and fills up the finite volume of field space with uniform density in the measure defined by the field metric, while the total energy density approaches the Planck scale.  It's clear that when the energy density gets to the Planck scale even the most hard core effective field theorist must admit that they have no idea what the actual state of the system is.
However, following effective field theory as far as it can take us, it is clear that the rapid time dependence of the zero mode in the crunching region of the CDL instanton will excite a high entropy state of the nonzero modes of the field theory.  When that entropy exceeds $A_{max}^{3/4}$ we need the real quantum theory of gravity to determine what is going on.  If that theory obeys the CEP, then unitarity and detailed balance tell us that the state will make a transition back to its dS progenitor, which has much more entropy.
 The Covariant Entropy Bound provides us with a completely finite measure of the probability of remaining in the crunching state, which is thermodynamic in nature.  It seems like a much more robust principle for deciding what the fate of this state of the system is, than some speculative claim that it is terminal.  The CEP predicts that for potentials above the Great Divide, the system will spend most of its time in the lowest dS minimum, despite the availability of other minima of lower "energy density". 

When the potential is below the Great Divide, we have a much less clear picture of what a possible theory of Quantum Gravity could be, which reproduces the properties of CDL transitions.  Let's consider the "decay of flat space", since the dS case introduces no new wrinkles.   Our remarks about the implausibility of the Big Crunch as a "terminal vacuum" remain unchanged in this case, but it is much less clear whether there is any sensible description of the time evolution of the system.  In ordinary QFT, when we find an instanton describing the decay of a false vacuum, it also describes the fate of local excitations of the vacuum.  If a bubble of true vacuum nucleates a distance $r$ from a local excitation, then that excitation is swallowed by the bubble in a time of order $r$ and decays into excitations of the true vacuum.   In a theory of gravity a generic localized excitation of Minkowski space is a black hole.   Suppose a Crunching bubble nucleates a distance $r$ from a black hole of radius $R$.   If $r \ll R$ the bubble will be smaller than the black hole when they collide and it will be swallowed by the black hole.
Since CDL bubbles expand in a causal manner, the bubble will be crunched inside the hole and will not effect things outside a radius of order $r + R$ from the hole.  Most localized excitations inside the horizon of the hole will not encounter the bubble. If $R \ll r$ it seems plausible that the horizon of the black hole will be swallowed by the bubble, but it will remain a black hole in the crunching space-time.  It's not clear that we know what the fate of such an object is.   What is clear is that the fate of an "unstable Minkowski space" due to CDL tunneling depends on the state of excitation of that space-time.   A dense enough collection of large black holes would postpone the onset of the instability until those black holes decayed.   All of this confusion may lead one to suspect that Lagrangians below the Great Divide, are part of what has become known as The Swampland.  They are low energy effective field theories, including gravity, which may not actually be realized by any consistent theory of quantum gravity.

\subsection{Transitions between dS space and a spacetime with vanishing c.c.}

There are actually three distinct types of effective Lagrangian that might have CDL instantons mediating transitions of this type.  In the first the potential has a stable minimum where it vanishes.  In the second two the potential goes to zero at infinity, either rapidly enough that there is no accelerated expansion, or less rapidly, as in quintessence models.   There are no plausible examples of the latter behavior in string theory\footnote{Recent bounds based on Swampland ideas allow for accelerated expansion, but there are no actual examples.} and no understanding of the nature of the quantum theory of gravity in quintessence spacetimes, so I will not consider that category of model. 

Let us instead focus on models of perturbative string theory with ${\cal N} = 1$ supersymmetry in four dimensions.  Many such models exist to all orders in perturbation theory and they all have moduli fields which, as a consequence of symmetries, can have no superpotential in their effective Lagrangian to all orders in perturbation theory.  Non-perturbative superpotentials are allowed by the symmetries, and are to be expected if we follow GellMann's Totalitarian Principle\footnote{Everything that is not forbidden is compulsory.} .  For known models, symmetry arguments show that the potentials fall too rapidly at infinity to have quintessent behavior.   In the case of the dilaton in weakly coupled heterotic string theory the potential of the canonically normalized field $\phi$ has the asymptotic form $e^{- e^{\phi L_S}}$, where $L_S^{-2} $ is the string tension.  Other moduli fields have potentials that fall at least exponentially.

First let's consider the case of an isolated zero energy minimum.  The case of a stable minimum at zero energy density corresponds to a supersymmetry preserving point that also preserves a (necessarily discrete abelian) R symmetry.  We also hypothesize the existence of a positive energy local minimum of the potential.
The observable of string theory in Minkowski space can be considered an infrared finite modification of an S-matrix.  The results of\cite{weinbergir} tell us that whatever the correct definition of that object is, the inclusive cross sections for scattering with unobserved soft graviton emission with some energy resolution, will have a familiar perturbative form at low energy.  We now want to consider two different processes: the formation by scattering of a localized meta-stable excitation with the fields near the false minimum over some region ${\cal R}$ and the CDL instanton for the decay of the metastable minimum.  Note that thinking of the latter process as a limit of dS to dS tunneling, we would guess that this is indeed a decay, because the backward tunneling rate vanishes in the limit that the energy of the lowest dS minimum goes to zero.  

The first point to make is that {\it these two processes cannot be part of the same model of quantum gravity}.  The S matrix is an observable in asymptotically flat space-time.  Every matrix element, including the ones which create a finite metastable region in the false minimum, exists in a Hilbert space with exact super-Poincare symmetry.  By contrast, the zero c.c. spacetime that evolves from the CDL decay has only the asymptotic symmetries of a 3-hyperboloid of constant negative curvature.  Moreover, despite some attempts\cite{frwcft}, we have no idea how to define the quantum theory or the observables in the CDL universe.   What is certain is that a definition in terms of some kind of S-matrix in the time symmetric extension of the expanding CDL bubble does not work.  If we take a generic state of incoming localized wave packets in this universe it will form a spacelike singularity that extends over an entire finite time Cauchy slice of the universe and does not have the future asymptotics of the original space-time\footnote{I suggested this based on an analogy with a similar criticism of an S-matrix proposal for dS space, but it was proven by\cite{boussofreivogel}.}.   

The phenomenon of different solutions of the same low energy effective Lagrangian not being part of the same model of quantum gravity is common in string theory.  We've already seen an example of it in our study of negative c.c. Big Crunches.  The AdS solutions at a stable negative value of the potential might belong to a well defined model of quantum gravity, but the crunching CDL instanton solutions of the same equations of motion certainly have nothing to do with that model.  A Lagrangian that is not in the Swampland, may have solutions that are.

The above paragraph implies that one cannot argue for the existence of a decaying dS space by constructing an effective potential with a meta-stable positive energy minimum to account for the behavior of scattering amplitudes in asymptotically flat space.  Indeed consider what happens when one prepares fields at infinity which will create a region of space of size $R$ in which the scalar field is constant and equal to its value at the positive energy minimum.  If $R$ is much smaller than the Hubble radius of the dS solution at that minimum, then gravitational back reaction is small and the excitation decays as one would expect in a field theory in Minkowski space.   In principle, we could map out the potential by doing such theoretical experiments\footnote{Non string theorists may want to note for further use below the fact that string theory has no object in it that plays the role of an effective Lagrangian or potential.  These quantities are derived by comparing S matrix elements computed by stringy methods, or CFT correlators in the AdS/CFT correspondence, to computations done with an effective action. The procedure is inherently approximate and does not lead to a non-perturbative definition of an effective action.}.  As we raise $R$ towards the Hubble radius, gravity intervenes in an inevitable way.  Because of the gradient energy in the walls separating the positive energy region from the region where the field is at the zero energy minimum, we find that at some point where $R$ is strictly less than the Hubble radius, a black hole forms with a radius larger than the Hubble radius.  This black hole then decays in the universal manner of all black holes of the same mass.  It decays back to jets of particles in asymptotically flat space, plus unobserved gravitational radiation below the energy cutoff.  It does not decay by the CDL process.  Both the rate of decay and the final state are radically different.  Typically, the instanton decay rate will be much much slower than the Hawking formulas suggest, and the whole black hole will disappear much more rapidly than CDL predict.  Although the black hole evaporation will take longer than the dS Hubble time, the region that one associated with the "local metastable dS space" will collapse to a singularity in a time of order the Schwarzschild radius, which is not much larger than the Hubble time.  The number of "e-folds of inflation" experienced by localized objects in the putative dS patch will be order one.  Thus, there is no sense in which creation of this patch with the local geometry of dS space is equivalent to the creation of a dS space that decays by instanton transitions.  It makes even less sense to imagine creating, inside of a theory of quantum gravity in Minkowski space\footnote{Similar remarks are true for AdS boundary conditions.}, a local region that evolves to an inflationary universe.   In an inflationary universe the causal diamond of a trajectory that was initially in an approximately dS patch, expands so that it is in causal contact with other inflationary patches.   A detector traveling along that trajectory detects the inflationary regions in its FRW past\footnote{These regions are on a single spacelike slice through the diamond.} as fluctuations in its microwave sky.  However, if the whole patch is contained in a black hole of radius not much larger than the inflationary Hubble radius, then such an expansion of the causal diamond is impossible.  Causal diamonds inside a black hole always have holographic screens whose area is no bigger than the black hole horizon area.   We conclude that reconstruction of a potential with a local positive energy minimum from low energy scattering data in a model of quantum gravity in Minkowski space, or from field theory correlators on the boundary of an asymptotically AdS space, does not tell us anything about another 
model, which might have a CDL transition from that minimum to a zero c.c. FRW or a negative c.c. Big Crunch.   Different solutions of the same low energy effective field theory are not states in the same model of quantum gravity.  One solution may be in the Swampland, while another is perfectly well defined. An effective action derived by approximate computations in a well defined model of quantum gravity, cannot claim to be a rigorous justification for other models, with different asymptotic boundary conditions.

It's rather amusing that these considerations lead to conclusions at odds with the seemingly similar arguments of\cite{dSswamp}.  Those papers try to put bounds on the possible behavior of potentials that could possibly derived from a consistent theory of quantum gravity.  The general idea is that there is something wrong with a dS space that has too long a lifetime.  There are two different ways in which the considerations of this paper lead to a completely different procedure.  Again, the best way to think about this is in terms of perturbative string models with $N = 1$ SUSY in four dimensions.  GellMann's totalitarian principle leads us to suspect that most of the moduli space of those models will belong to the swampland.  A generic point in moduli space will not lie at a stationary point of the non-perturbative superpotential that GellMann's principle leads us to expect.  Thus, the space-time associated with it will be some sort of cosmology, and therefore have a Big Crunch or Big Bang singularity which is inconsistent with the existence of the scattering amplitudes that we can in principle construct to all orders of string perturbation theory.  The fact that the string perturbation series is not Borel summable shows that there is no contradiction between these two conclusions.  Perturbative moduli space completely distorts the true nature of the class of consistent models.
\begin{figure}
\begin{center}
\includegraphics[scale=0.5]{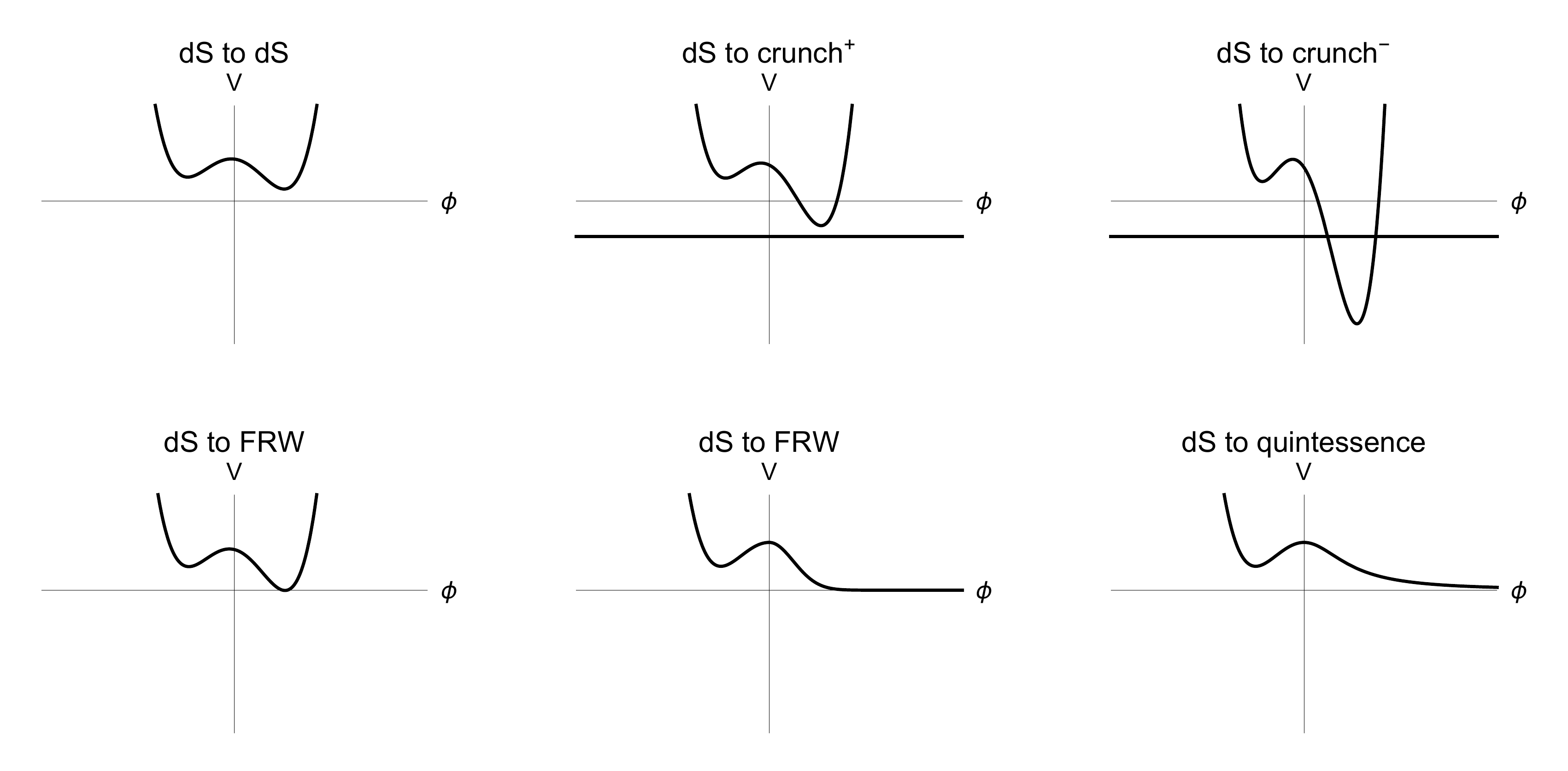}
\caption{Potentials for CDL instantons discussed in the text. Two cases, tunneling between two dS minima or between a dS minimum and a negative c.c. Crunch above the great divide can be interpreted as transitions in a finite quantum system between two ensembles of different free energy, with the transition rates computed from areas in the classical geometries.  In both cases, the lowest lying dS space is the configuration in which the system spends most of its time.  Transitions to a crunch below the great divide probably live in the swampland, as do transitions to quintessence.   Transitions to a zero c.c. minimum, or an asymptotic region of moduli space where the energy density asymptotes rapidly to zero, could describe actual decays of dS space, but we do not have a good idea about the quantum theory dual to such asymptotically FRW geometries.   It is certainly not the super-Poincare invariant scattering theory that would be defined at the zero c.c. minimum.  
}
\label{l}
\end{center}
\end{figure}

Let us however concentrate on a point in perturbative moduli space, which is a stationary point of the potential with a discrete R symmetry, which guarantees that the superpotential vanishes at the minimum.  The low energy effective field theory analysis leads us to expect a super-Poincare invariant model of quantum gravity.  Generically\cite{dineseiberg} we would not expect this to occur at weak coupling, so it's not obvious how to compute the S-matrix at this point from the string perturbation series, but there's a plausible argument that the model exists.  Let's also recall that models with $N = 1$ SUSY often contain asymptotically free gauge theories with chiral matter.   Non-perturbative physics in these models will generate superpotentials for gauge theory moduli with energy scales and typical ranges of field variation that are $\ll M_P$.  We know\cite{seiberg} that the non-perturbative superpotentials generated by gauge theory dynamics often have SUSY preserving vacua with unbroken R symmetry (typically only at discrete points in moduli spaces unless the number of R charge $0$ moduli exceeds the number of $R$ charge $2$ moduli).  

I confess to not having made a systematic study, but I would bet that among these plausibly consistent\footnote{Note that we've been using effective field theory only as advocated above, to find plausible consistent super-Poincare invariant models of quantum gravity.} models one can find examples where the gauge dynamics generates locally stable positive energy minima, with only tiny gravitational corrections.  The approach of \cite{dSswamp} has to declare that for no apparent reason, these models are in the Swampland, because they have meta-stable "dS" points.  By way of contrast, the approach presented in this note views these minima as metastable states with negligible gravitational back reaction as long as the region of space excited into the false minimum is small compared to the "dS" Hubble radius.  For larger values of the region one instead finds a black hole, which decays back into jets of particle excitations of the super-Poincare invariant QG model.   The existence of the meta-stable minimum does not provide evidence for or against the existence of a model of quantum gravity in which the dS space decays via the CDL instanton.  It's entirely reasonable to guess that that solution is in the Swampland, but this says nothing about the consistency of the model whose low energy physics is modeled by the effective field theory at the super-Poincare invariant point in moduli space.

The situation for CDL tunneling from dS space is summarized in Figure 2.  Two cases, tunneling between two dS minima or between a dS minimum and a negative c.c. Crunch above the great divide can be interpreted as transitions in a finite quantum system between two ensembles of different free energy, with the transition rates computed from areas in the classical geometries.  In both cases, the lowest lying dS space is the configuration in which the system spends most of its time.  Transitions to a crunch below the great divide probably live in the swampland, as do transitions to quintessence.   Transitions to a zero c.c. minimum, or an asymptotic region of moduli space where the energy density asymptotes rapidly to zero, could describe actual decays of dS space, but we do not have a good idea about the quantum theory dual to such asymptotically FRW geometries.   It is certainly not the super-Poincare invariant scattering theory that would be defined at the zero c.c. minimum.  

\subsection{Transitions from AdS space to negative c.c. Big Crunches}

This subject has been investigated extensively\cite{barbrabharlowherthormalda} and I invite my readers to consult the excellent references.  The bottom line is that most examples of instantons for these decays make no sense in the AdS/CFT correspondence.  They correspond to perturbations of a CFT by {\it irrelevant} operators.  There is a small class of cases for which a dual field theory interpretation makes sense\cite{hhmalda}.   These occur in AdS spaces whose classical solution sits at a tachyonic maximum that is allowed by the Breitenlohner-Freedman stability bound.  The dual CFT is supersymmetric and stable and is $2 + 1$ dimensional. In the dual CFT there is a SUSY violating relevant perturbation, of the schematic form $\phi_1^2 - \phi_2^2$ which would be unbounded from below in flat space.  However the CFT dual of global AdS space (as opposed to the Poincare patch of AdS space that represents the near horizon limit of a stack of infinite black branes) lives on $R \times S^{2} $, with the radius of the sphere equal to the AdS radius of curvature.  If the coefficient of the unbounded relevant operator is sufficiently small, the field theory is stable but its bulk dual is the instanton going from AdS to the Crunch.  Parenthetically, we should note that in the original Hertog-Horowitz discussion of this system, the instanton was interpreted as a perturbation by the {\it cube} of the relevant operator, which is marginal to leading order in $1/N$.
In this case one has to invoke a {\it dangerous irrelevant operator} to stabilize the system.  Maldacena pointed out that the instanton solution satisfied {\it both} the boundary conditions corresponding to the relevant perturbation and to its cube.  The question of which interpretation one is using corresponds to a choice of the boundary conditions on the fluctuations around the instanton.  Thus, the CDL instanton alone does not indicate which boundary field theory describes the physics.  Finally, note that the instanton does not have the interpretation of a transition in the boundary field theory.  Rather, in the most transparent case where the perturbation is relevant, it is a renormalization group flow from a stable CFT to a massive field theory that is unbounded from below in flat space but stable on the sphere. Since the theory must live on the sphere, the RG flow never actually completes, because there is an IR cutoff.

\subsection{Discussion}

We have seen that a field theory with false vacuum decay in flat space has a BD state in dS space, which describes thermal transitions back and forth between the multiple minima of the potential, as long as the potential is above the Great Divide and has no locally stable stationary points with exactly zero energy.  The predictions for observables in any causal diamond of the space-time are finite and unambiguous.  The "God's Eye" observables that would test the predictions of EI for relative frequencies of minima on a space like slice at infinite global time correspond to states orthogonal to all finite norm states in the BD Hilbert space.  The predictions for them are ambiguous and depend on the space like slice.
In the limit of infinite time, they do not correspond to normalizable states in the BD Hilbert space and so one cannot make unambigious predictions.

CDL gravitational instantons for dS to dS transitions give support to this picture and identify the back and forth transitions when gravity is taken into account, as transitions in a finite entropy system.   For the case of transitions to a negative c.c. crunch, low energy models fall into two classes, depending on whether the scalar potential in the CDL effective field theory is above or below the Great Divide.  Above the Great Divide, a plausible interpretation is again a transition in a finite entropy system, with the inverse transition rate determined by the principle of detailed balance. In this interpretation the entropy of the Big Crunch region is defined by the size of its maximal causal diamond.  Models below the Great Divide have no clear quantum interpretation and may be part of the Swampland. They do not resemble decays, because excited black hole states of the "false vacuum" can suppress the transition.  In neither case is there support for the idea that the Big Crunch is a "terminal vacuum".  The classical solutions for the scalar field in the crunch region do not remain in the basin of attraction of the negative c.c. minimum.   

For locally stable AdS solutions, the AdS/CFT correspondence shows that CDL instantons for decays to a crunch, do not correspond to states in the CFT but perturbations by either irrelevant operators (in which case they are ill defined) or by relevant operators that are unbounded from below, but can be stabilized for small enough perturbation, by the curvature of the boundary sphere.  

The overwhelming message is that field theory in dS space, a careful analysis of CDL instantons, and the AdS/CFT correspondence show no positive evidence for the phenomenon of false vacuum Eternal Inflation and suggest strongly that it is not a correct description of any model of quantum gravity.  We've also learned the important lesson that different solutions of a low energy effective field theory do not all correspond to the same model of quantum gravity.  In particular, establishing the existence of a positive energy minimum in a low energy field theory that matches the S matrix or boundary correlators of a string theory model of gravity in Minkowski or AdS space {\it does not say anything} about the existence of a metastable dS space that decays via a CDL instanton, and is inconsistent with an inflationary universe with causal diamonds much larger than the dS Hubble radius "inside a local region".  

\section{Landscapism and Landskepticism}

In\cite{isolated} I argued that in a gravitational model whose low energy effective field theory had a scalar potential with two isolated zero energy minima, these minima did not correspond to quantum states in the same model.  The argument is essentially identical to that we gave above for a metastable  positive energy minimum.  In non-gravitational field theory we would establish the claim that the minima corresponded to states in the same model by creating an arbitrarily large region of constant field in one minimum starting from scattering data around the other.  In models of gravity, the size of the "alien" region is limited by black hole creation, and the black holes decay back to excitations of the minimum in which they were formed.  When I presented this result in a group meeting at Rutgers in 2000, it was pointed out that the same thing would happen if one tried to create a region too far away in a moduli space of string models.   The issue is the energy in the domain wall between the two different values of the field.  I mentioned this extension of the result at the 2004 Strings meeting.   In 2008 Nicolis discovered this result independently\cite{nicolis}. 

String theorists are prone to saying that the coupling constants of the theory are expectation values of fields, implying that different values for these constants are states of the same model, in much the same way that, in non-gravitational quantum field theory, different asymptotic minima of the scalar potential represent different states of the same underlying Hamiltonian.   In fact however, in every non-perturbative approach to string theory, these values are parameters rather than expectation values, in agreement with our intuitive argument based on black holes.  Matrix Theory\cite{bfss} is a non-perturbative definition of models of quantum gravity with a super-Poincare group having at least 16 real supercharges.  The set of all such models has a multidimensional manifold of moduli, which are realized by compactifying $N =4$ super Yang Mills theory on different manifolds, and taking the size of the gauge group to infinity.  Compactifications of SYM on tori with boundary conditions preserving some supersymmetry is dual to 11D SUGRA compactified on T-dual tori. Since the quantum dynamics of SYM theory does not change the manifold, these are all different models rather than different states in a single model\footnote{Parenthetically, one may also note that these models also illustrate the importance of SUSY in quantum models of gravity in Minkowski space.  We can compactify SYM theory on manifolds that have no Killing spinors, leaving some non-compact transverse coordinates.  The "soft" violation of SUSY lifts the moduli space of non-compact coordinates and the model no longer has a scattering matrix at all.}.

Similarly, in AdS/CFT, the quantum dynamics of M-theory on a variety of manifolds of the form $AdS_D \times {\cal K}$ is defined by a CFT.  Continuous families of such CFTs are {\it conformal manifolds} : the space of exactly marginal perturbations of a single CFT in the family.  The complex coupling of ${\cal N} = 4$ SYM theory, which we mentioned above, is one example of this phenomenon, and there are many others.
Parenthetically, AdS/CFT also gives us many examples where different solutions of the same bulk supergravity Lagrangian correspond to different field theory models.  In particular, the value of the c.c. is always discrete in Planck units, and related to the asymptotic spectrum of high dimension operators.
Furthermore, when the gravity dual has a scalar potential such that the AdS state is identified as a particular stationary point, FRW models which start with generic boundary conditions at other points on the potential, are completed unconnected to the CFT and probably represent models that are in the Swampland.

There is yet another argument that the continuous parameters that appear in many string theory models of quantum gravity should not be thought of as "vacua" of a single Hamiltonian.  In QFT, scattering amplitudes in the high energy fixed angle limit probe the UV fixed point that defines the model.  The differences between different vacuum states become negligible in this limit.   In models of quantum gravity, the fixed angle high energy limit is deep inside the kinematic region that is dominated by the production of black holes\cite{penroseetal}.  Black holes of large energy have very low Hawking temperatures in Minkowski space and their decay processes are sensitive to the properties of the lightest particles in the theory.   We thus find different answers for high energy fixed angle scattering at different points in moduli space, again suggesting that these are not different infrared limits of a single UV Hamiltonian.

In the previous section, we mentioned that string theory gives us many examples of perturbative models in Minkowski space, which have ${\cal N} = 1$ super Poincare invariance in four dimensions, and a perturbatively unitary S matrix to all orders in perturbation theory\footnote{Some caveats to this last statement are that the S matrix has infrared divergences and that one does not have an offshell formalism to deal with the instability of most of the higher mass modes of the string.  Inclusive cross sections will be finite, and Sen\cite{sen} has made some progress on dealing with the fact that most of the particles in the tree level S-matrix are only resonances at any finite coupling.} .  We also said that symmetries forbid the appearance of a superpotential in the low energy effective Lagrangian to all orders in perturbation theory, but speculated that in most cases, there would be non-perturbative corrections to the superpotential.
Now that we understand that the continuous moduli of these models do not represent different states of a single model, what would such a superpotential signify?

A clue is provided by the Fischler-Susskind\cite{fsm} mechanism for tachyon free SUSY violating perturbative string models in Minkowski space.  Fischler and Susskind showed that such models suffer from a divergence in one loop string perturbation theory, which can be formally canceled by making a "small" perturbation of the background dilaton and gravitational fields.  The perturbation is said to be "small" only because it cancels a one loop effect. The idea is that the divergence comes from a tadpole for the dilaton field, which suggests a "shift in the background classical solution". Unfortunately, the perturbations are not really small for all times.  The famous Hawking-Penrose theorem shows that any solution of the low energy field equations starting on the slope of a potential with no positive c.c. minima for the dilaton field is a singular cosmology, either a Big Bang or Big Crunch.   The "small perturbation" is so drastic that we cannot even define the asymptotic states of the perturbed model.  

 Note that even a locally stable AdS or dS solution is a huge perturbation of the asymptotic region of space-time where the S matrix is defined\footnote{We'll discuss the special case of a very small negative c.c. below.}.   Thus, the Fischler-Susskind mechanism suggests that tachyon free SUSY violating string theories are not valid models of gravity in Minkowski space-time and are not amenable to perturbative analysis.   It may well be that, like superstrings with gauge anomalies, they are part of the Swampland.  

Returning to models with SUSY and perturbatively vanishing superpotentials, our hypothetical non-perturbative superpotentials are non-perturbative analogs of the FS mechanism.  That is, most of the points on moduli space do not correspond to models of quantum gravity in Minkowski space, and may not be consistent models at all.  Certainly the effective field theory of the generic model is a cosmology and one must understand how to deal with the inevitable cosmological singularities before declaring that it is sensible.  Current methods of string theory give us no clue about how to deal with such situations.  If there are stationary points of the potential with small negative or vanishing value of the c.c. these might correspond to valid models in AdS or Minkowski space. We'll argue below that these models are exactly supersymmetric.  

It's important to realize that the entire procedure just outlined for finding (meta) stable AdS minima of a non-perturbative effective potential  is purely hypothetical and has no basis in well founded string theory calculations.  String theory consists of a collection of models in asymptotically flat or AdS space for which one can calculate boundary correlation functions and argue that in certain limits the same calculations match those of a bulk effective field theory including gravity.   There is no definition in the setup of the boundary theory that corresponds to an off shell effective potential, no bulk "path integral" from which one can integrate out UV degrees of freedom.  We do not understand how quantum field theory emerges from these theories of quantum gravity or what the limits on its formalism are.  The considerations of \cite{ckn} show us that no standard cutoff procedure captures those restrictions.  The failures of field theory to match correct predictions of models of quantum gravity are not just ultraviolet in nature.

The hypothesis of the String Landscape is entirely based on low energy effective field theory ideas about finding "vacua" by minimizing an effective potential.  Everything that's been said above indicates that this idea has no validity in genuine models of quantum gravity.  The use of low energy effective field theory should instead be thought of as a consistency condition.  If one has a quantum gravity model with a set of well defined correlation functions, it will have a low energy sector, including gravitons and other light particles (in AdS/CFT, "light particles" is replaced by low dimension primary operators).   General principles\cite{weinberg}\cite{polchheems} then assure us (in the case of AdS/CFT we must also assume an AdS radius large compared to all microscopic length scales) that some correlation functions can be approximately calculated in a low energy effective field theory, which has a solution corresponding to the assumed asymptotic spacetime of the quantum gravity model.  If in addition the model has some exact symmetries, which guarantee stability of the asymptotic solution against corrections to the low energy effective field theory, then we can be confident that we have a valid model.   In all known cases the exact symmetry includes supersymmetry.  No one has ever found a consistent model of quantum gravity in Minkowski space of more than $1 +1$ dimensions, with broken SUSY.  All known AdS/CFT dualities with AdS radius much larger than microscopic scales are relevant perturbations of exactly supersymmetric conformal field theories.  I'll review below the most convincing argument that SUSY is necessary in large radius AdS models.

Many proposals\cite{kachru} for the Landscape of "realistic" string theory models start by finding a plausible non-perturbative correction to the superpotential of a super-Poincare invariant model with four supercharges in a space-time manifold that is conformal to\footnote{The conformal factor depends only on the coordinates of the compact manifold ${\cal K}$.} the form $M^{1+3} \times {\cal K}$, with compact ${\cal K}$.  The superpotential has a supersymmetric AdS solution with small superpotential, and thus small negative c.c. . This plausible non-perturbative superpotential does not correspond to an actual construction of the CFT that would define such a model, but at best is a calculation done on a system of D-branes embedded in flat space-time with dimension $10$, which one then assumes can be compactified down to four dimensions.

 While these constructions are not rigorously justified they are plausible.  The most serious issue, in my opinion, is the contention that one can make the AdS radius much larger than the size of the compact manifold.  All well established examples of large radius AdS/CFT have a compact manifold of dimension $2$ or greater whose radius is comparable to that of the AdS space.  In Appendix A we'll present an argument based on the properties of AdS black holes, that this is in fact necessary.

The next step in the construction of "realistic" models involves "adding an anti-brane to break supersymmetry and make the c.c. positive".  This is supposed to be a small modification of the model, calculable in low energy effective field theory, and that seems manifestly incorrect.  Whatever the detailed properties of the AdS model, we know that it is a conformal field theory with an infinite spectrum of stable states at arbitrarily high energy.  dS space cannot support such a spectrum.  Even if one ignores the contention that it has finite total entropy, it is a simple fact about solutions of Einstein's equations that its black hole spectrum is bounded.  A small low energy correction to the Hamiltonian is supposed to completely eliminate an infinite high energy spectrum of stable eigenstates.  

Disregarding this flagrant problem with the program of constructing "reliable" dS models "in string theory", let's consider what this procedure says about the Landscape of such models.   So we now consider a different point in the moduli space of perturbative super-Poincare invariant models where we can plausibly find another large radius supersymmetric AdS solution of the equations of motion of a non-perturbative effective superpotential.   According to the rules of the AdS/CFT correspondence, this is a different conformal field theory, with a different spectrum of operators, even at high dimension.  Even if both lifted dS models exist, there is no argument that they are part of the same model and that transitions between them can occur.  The ideology of the Landscape is that models with very small negative c.c. are very rare in the Landscape, and this is certainly the case if one follows the logic of Bousso and Polchinski\cite{bp} as to how a negative c.c. is achieved.  Thus the minima of the potential on moduli space corresponding to these models are not close together and there is no argument that the two CFTs are related to each other.

The most sophisticated attempts to construct a set of boundary correlation functions which might provide a mathematical definition of the Landscape of String Theory goes by the name FRW/CFT\cite{frwcft}.   I think it is fair to say that the authors of those papers have not shown how to construct a Hilbert space interpretation of their calculations, and the program seems to have been abandoned some years ago.  However, there is something in the basic setup that echoes our arguments that different dS minima in the Landscape are not part of the same model, even if the Landscape exists.   The key actor in FRW/CFT is a non-singular FRW space-time {\it obtained from the analytic continuation of the CDL instanton for decay of a particular dS minimum to one of the asymptotic regions of moduli space where "SUSY is restored"}.  These, along with analytic continuations of instantons for decay of flat space with SUSY violating boundary conditions on a torus\cite{evaetal}, are among the only non-singular FRW models.  The boundary conditions on allowed perturbations of this space time that could define boundary correlation functions are not known.  General outgoing low energy wave packets of massless fields on the future conformal boundary can be the time reverse of incoming wave packets that would have formed black holes of arbitrarily large size in Minkowski space.   In the CDL instanton space-time, these boundary conditions lead to a space-like singularity spanning the entire space, which intervenes between asymptopia and the time symmetric point of the Lorentzian CDL instanton.  This intuitive argument (which is my own) was turned into a rigorous result in GR in\cite{boussofreivogel}.  In addition, the authors of \cite{frwcft} argue that the boundary of spacetime is in fact fluctuating and that the correct observables are in fact those of a boundary Liouville quantum field theory, viewed not as a CFT but as a model of two dimensional quantum gravity.

Clearly, the classical construction here depends on the choice of dS space from which the CDL instanton tunnels.  I believe that one is supposed to argue that averaging over boundary Liouville fluctuations somehow takes all other CDL instantons in the hypothetical Landscape into account, but I have never understood how such an argument would work.  Again, even if one believes that the construction of meta-stable dS models is reliable, there is no clear argument about what the proper observables of the model are nor that different dS constructions are part of the same model.  Neither is there an interpretation of these correlators as transition amplitudes in a quantum mechanical model.

The conclusion that effective field theorists should draw from this is that unlike supersymmetric string models in flat or AdS space-time, many of which have at least perturbative definitions as mathematical models obeying the axioms of quantum mechanics, all literature on the String Landscape is speculation based on the unfounded notion that all string models with a given amount of SUSY are part of one single model and that it makes sense to define an effective action that encompasses all string models.  Every single non-perturbative construction of string models contradicts this claim: AdS/CFT, Matrix Theory, and $1 + 1$ dimensional models of various types, are all unitary quantum theories, which describe only a single point in moduli space.  Furthermore, arguments involving black hole formation prevent one from performing the usual manipulations one would go through to prove that two different solutions of an effective field theory are part of the same model, whenever the solutions have different asymptotic boundary conditions.  

How should effective field theorists react to these conclusions?   Some may wish to discard wisdom gleaned from string theory altogether, imagining that it is just a failed attempt among many to construct a theory of quantum gravity.  This attitude seems indefensible, for a field theorist who indulges in wild speculations about EI. EI was an attempt to guess a properties of a model of quantum gravity, based on the assumption that effective quantum field theory could be trusted whenever local curvature invariants were small.  We've known since the work of\cite{ckn} that this is not correct.  So if we reject string theory as a reliable guide to the construction of theories of quantum gravity, how can we take seriously models that make extreme speculations about what such a model would look like? String theory is the only successful attempt to construct mathematical quantum mechanical models whose properties correspond to many features one can derive independently by solving the equations of general relativity, coupled to other fields, and it has given us valid new ideas about many old subjects\footnote{Perhaps the most striking is the relation between non-Abelian gauge theories/chiral fermions and singular geometries of extra dimensions. }.

A more sensible attitude, which I share, is to accept that string theory defines some models of quantum gravity, but obviously not the one that corresponds to the real world.
One might still want to speculate that the idea of EI and a landscape, while not validated by anything but speculation in existing string theories, might be valid in the real theory of the real world.  One would aver that extant string theories, if the above arguments are correct, are just a small corner of all possible models of quantum gravity, and don't exhibit all of the rich possibilities that are exhibited by the EI theory of our world.  The first section of this essay was meant as a rebuttal to the latter point of view.

\section{Anthropic/Environmental Selection Arguments, Fine Tuning, and the Role of SUSY}

It's clear that the reason effective field theorists have started taking environmental selection arguments (and consequently EI) seriously is the failure, over a period of $50$ years, to explain the small size or vanishing of the c.c. by symmetry arguments that did not violently contradict other experimental data.  As the author of one of the first three papers to try to explain the small size of the c.c. in terms of the anthropic principle and the dynamics of scalar fields\cite{davieslindebanks}\footnote{The general idea of explaining the value of the c.c. by using the anthropic principle, but without any actual model, is due to Barrow and Tipler\cite{BT}.} I certainly have sympathy with this point of view. In that paper I made a guess that the anthropic bound was close to the observational bound. Later, Weinberg\cite{weinbergcc} made a much more careful and more celebrated, estimate, which showed that one only gets close to the observational bound by using the value of primordial density fluctuations measured by the CMB.  In a model where both of these parameters vary, one gets much worse results.  One of the great things about Weinberg's argument is that it is really {\it galactothropic} and therefore makes very few assumptions about the details of particle physics, chemistry, or exobiology\footnote{It probably fails however if Fred Hoyle's suggestion\cite{BlackCloud} that astrophysical gas clouds could be intelligent, and that life on planets is anomalous, is true.}.
Attempts to fix any but the most primitive parameters that determine the structure of the universe rely on many more assumptions, because we know so little about how to derive the existence of life, let alone intelligence, from physics.   It's silly to even think that we could come to a conclusion about such issues if we varied the gauge group or representation contents or Higgs structure of the standard model in random ways.

More importantly, most environmental selection arguments based on actual models do nothing but explain mysteriously tuned numbers in our current model of the world.  They do not lead to other checks of their validity.  Thus, in using environmental selection in theoretical physics we should adopt Andy Albrecht's Principle: "The physicist who uses the fewest environmental selection arguments to explain the world, wins."  

Despite this, the failure of the LHC to find any hint of SUSY or other proposed mechanisms for explaining the hierarchy between the electroweak and unification scales, has led to speculation about anthropic determination of the electroweak scale.
Most of these arguments suffer from multiple degeneracy problems, the relevant physics depends not only on the electroweak scale but on the tiny electron Yukawa coupling whose explanation is no less mysterious because we have a variety of models to explain its size using symmetries.  All of the arguments based on atomic and stellar physics give much less impressive results if we allow both the electroweak scale and the electron mass to vary randomly.  Furthermore, since neither atomic nor stellar physics is sensitive to details of the weak interaction, these arguments should probably be made in the context of the the Fermi theory of weak interactions supplemented by a neutral current self interaction and would allow many more variations of standard model parameters, including the existence of extra gauge bosons in regions where they're experimentally excluded.  Perhaps even "weakless"\cite{weakless} models give us anthropically acceptable models.  Models with low energy SUSY or technicolor certainly do, but are ruled out by experimental data to which anthropic arguments are not sensitive.

One's attitude to the little hierarchy problem is conditioned by one's attitude toward SUSY.  I have been impressed (some will say overly impressed) by the failure of all attempts to break SUSY in string theory models in Minkowski space.  This led me to conjecture long ago that families of CFTs corresponding to AdS spaces with curvature radius that could be dialed  larger than any microscopic scale, would all be supersymmetric.  Most of the data on AdS/CFT models confirms this\footnote{There are $3$ possible classes of exception known.  They are discussed in\cite{tbov}.}.  The paper of\cite{ov17}, known as the AdS Swampland conjecture, gave arguments to this effect.  In trying to understand those arguments, I realized that there was a more modest statement that was a rigorous mathematical conjecture in flat space string theory\cite{tbov}.  The original $AdS_5 \times S^5$ model was discovered by examining the near horizon limit of stacks of large numbers of parallel branes in flat space.  Most other models can also be realized as near horizon limits of stacks of branes in more complicated spaces.  The existence of limits of large numbers of parallel branes, whose gravitational backreaction gives rise to the large radius AdS geometry, is due to supersymmetry.  Stable, but non-supersymmetric configurations of branes exist, but are classified by small finite groups, called K-theory groups\cite{wittenhoravasen}, or generalizations of these to M-theory\cite{diaconescu}, and central extensions of these groups\cite{moorefreedsegal}.   The conjecture is that there is a maximal (probably modest) size to the generalized K-theory groups that arise in string theory.  As a consequence, there will be no large radius SUSY violating near horizon AdS geometries, because the brane stacks are small.

At the Festschrift for Lenny Susskind at Stanford in 2000, I put together the absence of SUSY breaking in Minkowski space, with the strong evidence that the c.c. is positive in our universe and conjectured that the breaking of SUSY was connected to the positivity of the c.c..  That was the year that I began to think of the c.c. as a parameter characterizing the large proper time and consequently (see below) high energy behavior of models of quantum gravity, rather than an energy density to be calculated in terms of more basic parameters.  The evidence that this was so, coming from AdS/CFT, may have influenced that change of attitude, but those facts had not sunk in on a conscious level.  In $2002$ I came up with a "hand waving at the horizon" argument for the connection.   The argument is simple to state, somewhat harder to justify in any sort of rigorous fashion.  The gravitino mass violates any R symmetry, a symmetry that does not commute with the supercharges.  $R$ symmetries are chiral.  R symmetry is also violated by the value, $W_0$, of the superpotential at the minimum of the potential.   In order to have a small c.c. in the presence of SUSY breaking, $W_0$ must be non-zero and of order $F m_P$.  The gravitino mass is of order $F/m_P$. Thus, SUSY breaking with small c.c. implies a gravitino mass and R symmetry breaking.  If SUSY breaking is, as I conjectured, connected to the finite cosmological horizon, then the dominant contribution to R symmetry violation must come from exchange of the lightest particle carrying the chiral R symmetry with the horizon.  Photons and gravitons are R neutral.   Assume then that the lightest R charged particle is the gravitino.   $R$ violating couplings will then be suppressed by $e^{- 2 m_{3/2} R_{dS}}$  However, the horizon has a huge number of low energy ($\sim R_{dS}^{-1}$) states, of order $e^{\pi (R_{dS} M_p)^2}$.  The absorption and re-emission amplitude will be, by standard second order perturbation theory, a sum of positive terms coming from the states with which the gravitino interacts.  Since it is a massive particle it can stay on the horizon for a proper time of order $\tau \sim m_{3/2}^{-1}$ .  Its interaction with the Planck scale q-bits on the horizon should make it perform a random walk with Planck scale time step, so it will cover an area  $c (m_{3/2}M_P)^{-1}$  .   Now consider the limit $R_{dS} \rightarrow\infty$.  $m_{3/2}$ should go to zero in this limit so the R violating term which will act to induce a gravitino mass in the low energy Lagrangian must go to zero.   We have just estimated it to be of order \begin{equation} \delta {\cal L}_{\slash{R}} \sim e^{ - 2 m_{3/2} R_S} e^{c (M_P m_{3/2})^{-1}} .  \end{equation} If $m_{3/2} $ has a leading scaling behavior $M_P (R_{dS} M_P)^{- a}$ then the right hand side of this equation blows up exponentially if $a > 1/2$ and falls exponentially if $a > 1/2$.   Thus, the only consistent power law behavior is $ a = 1/2$.  Subleading corrections given by a multiplicative factor in the mass formula of the type $ (1 - a (R_{dS} M_P)^{-1/2} {\rm ln} [R_{dS} M_P ] )$ then make the formula consistent if
\begin{equation} m_{3/2} = \sqrt{2c} M_P (R_{dS} M_P)^{-1/2} = \sqrt{2c} 10^{- 11.5} {\rm GeV} = \sqrt{20c} 10^{-3} {\rm eV} . \end{equation} This gives typical SUSY mass splittings of order \begin{equation} \Delta m \sim (80 c)^{1/4} TeV . \end{equation} Thus, with plausible values for $c$ this idea predicts SUSY partner masses compatible with but close to current lower bounds.  A more detailed model\cite{pyramid}, motivated by making this SUSY spectrum compatible with gauge coupling unification, predicts Dirac contributions to the gluino mass, which make it heavier than squarks by a factor $\sim \sqrt{2\pi / \alpha_3}$.  The main problem with the model is the little hierarchy problem.   Note that the very light gravitino is compatible with cosmological bounds but the model does not provide a WIMP candidate, compatible with the failure of WIMP searches.  The latter failure has contributed to the community's disenchantment with SUSY.  In the models of\cite{pyramid} the dark matter is a baryon of a new strongly interacting SUSY gauge theory.   It might also be (see below) composed of tiny primordial black holes.

If SUSY breaking is tied to the c.c. then the environmental selection criteria for the c.c. are completely changed.   The c.c. affects the QCD scale, the value of the fine structure constant, and assuming one finds a solution to the little hierarchy problem in this framework, it might affect the weak scale as well.  It's not impossible then that the c.c. will need to be fixed to its correct value in order to get particle physics right.  In that case Weinberg's bound should be viewed as a constraint on the product of the cube of the density contrast and the dark matter density at the beginning of galaxy formation.  In the penultimate section of this essay we'll see that in the model of inflation that arises naturally when thinking of the Covariant Entropy Bound as a fundamental principle, the primordial density contrast seems to be an undetermined parameter (apart from a constraint that it be small), while the dark matter density is likely to be computable.  If we view $\rho_D$ and $\Lambda$ as being fixed by computation and by particle physics forms of environmental selection, then Weinberg's bound becomes a bound for the density contrast $Q$.  It takes the form
\begin{equation} Q \geq  (\Lambda/\rho_D)^{1/3} . \end{equation}  The right hand side is quite close to the observed value of $Q$, and we should also note that the theoretical model only makes sense if $Q$ is small.  Thus we have {\it a priori} bounds $1 \gg Q \geq  (\Lambda/\rho_D)^{1/3}$.   Also, as we'll see, changing the value of $Q$ changes the reheat temperature of the universe, as well as baryogenesis.   

The point of reviewing this somewhat obscure idea about SUSY breaking is to point out that a shift in theoretical framework can completely change the nature of anthropic argumentation.  It's similar to the point we made about switching between the standard model and effective field theory at the scale of nuclear physics in discussing the anthropic constraints from the existence of atoms and stars.  Anthropic arguments, by their nature, do not depend on much physics beyond the scale of nuclei.  The cosmological/galactothrophic arguments are similarly describable by an effective classical model with a few parameters.  Changing to a more UV complete description can change their nature dramatically, but when one bases that UV complete description on non-anthropic data, there's a bit of cheating going on.  One is not saying "What does the model have to do in order to sustain life as we know it?", but what must it do to sustain life and also explain a lot of other experimental data that is irrelevant to life. Thus, one is putting together two rather different kinds of arguments.  For example, we have a perfectly good model called technicolor, which has no fine tuning problem for the Higgs and gives us a perfectly livable universe with atoms and stars.  Its only problem is that it doesn't agree with other data, to which anthropics are not sensitive.

The use of the hypothesis of Cosmological SUSY breaking to modify anthropic arguments has a somewhat different flavor.  It uses a shocking theoretical guess connecting two different scales that characterize our world, makes predictions that are on the edge of being ruled out, and also changes the nature of anthropic bounds in a dramatic way.

My own conclusion from all this is that the mechanism for and scale of SUSY breaking are the most important problems at the interface of phenomenology and the attempt to construct models of quantum gravity.  Cosmological SUSY breaking answers these questions in an interesting way, which has consequences for experiments that will be done in the next few decades.  Its major problem is the little hierarchy problem, but existing low energy models, called the Pyramid Schemes, already contain elements, additional chiral singlets and Dirac contributions to gaugino masses, that have been invoked to ameliorate the little hierarchy problem.  The interesting Twin Higgs mechanism, in which an exact discrete symmetry forces the Higgs to be a pseudo Goldstone boson of an accidental continuous symmetry has not yet been explored in the Pyramid Scheme context.  It should be.

\section{Why the c.c. is Not an Energy Density}

This section was my talk at the workshop, so I will be brief.  The principal evidence for this contention is the vast array of well established AdS/CFT correspondences.  For a CFT on $R \times S^d$, with a sphere of radius $R_{AdS}$,  the entropy at high temperature is \begin{equation} S(E) = 
c_T A T^d  , \end{equation} where $A$ is the area of the sphere and $c_T$ a dimensionless constant characteristic of the theory.  With $T = (dS/dE)^{-1}$, this gives the log of the density of states at high energy, as a function of $E$. If $c_T\gg 1$ {\it and} there is a large gap in the energy spectrum below which the density of "single particle" primary states grows only like a power of the energy, then the model has an approximate dual description as quantum gravity in a space-time $AdS_{d + 2} \times {\cal K}$\footnote{The arguments of Appendix A suggest that the linear size of the compact manifold ${\cal K}$ is of order $R_{AdS}$.}  .  $c_T$ can then be calculated from the black hole entropy formula and is of order $(R_{AdS}/L_{d+2})^{\frac{d - 2}{d - 1}}$ where $L_{d + 2}$ is the Planck length in AdS space.  Thus, the AdS radius is a property of the high energy spectrum of the model.  Almost all known models with the required gap (exceptions are described in \cite{tbov}) are superconformal field theories.  For these, the c.c. is protected against renormalization in the bulk dual theory, by SUSY.  However, we can perturb them by SUSY violating relevant operators, important only at low energy.  Conventional effective field theory reasoning would lead us to expect a calculable low energy contribution to the c.c. .  This is not what happens.  Instead, the bulk image of the RG flow ignited by the relevant perturbation is an asymptotically AdS spacetime which violates the AdS isometries, leaving only rotation invariance.  The asymptotic space-time and the value of the c.c. are unaffected.   The actual field configuration does not resemble AdS space with a different c.c..  In the region where it differs from asymptotic AdS, it does not have the isometries of AdS.

We should note that if we followed these RG flows for CFT on flat space we would often find singular bulk geometries with timelike singularities. These are associated with IR physics of the gapped endpoint of the flow. Occasionally there are bulk flows to other AdS geometries. Most of those have tachyonic instabilities already in the "consistent truncation" of the SUGRA solutions that neglects Kaluza Klein modes on ${\cal K}$.  There is only one example where a convincing argument has been given that there are no such tachyons\cite{perlmutter} at all, even in the Kaluza-Klein sectors.  There are two points to be made about all these examples.  If the relevant coupling is small in AdS units, then the theory on the sphere, which describes the global AdS geometry, is insensitive to the new "fixed point" or gapped theory, because the IR behavior is cut off.  Secondly, even in cases where there {\it might} be a new description of the IR of the field theory in flat space in terms of a non-SUSY AdS space with large radius, the new c.c. has nothing to do with loop corrections to the bulk Lagrangian.   

A more philosophical, but IMHO deeper, reason to believe that the c.c. is not an energy density, is Jacobson's derivation\cite{ted} of Einstein's equations as the hydrodynamic equations of the entropy law $S = A/4 A_P$ for all causal diamonds in any Lorentzian space-time.  That derivation leads to
\begin{equation} k^{\mu} (x) k^{\nu} (x) (R_{\mu\nu} - \frac{1}{2} g_{\mu\nu} R - \kappa^2  T_{\mu\nu} ) (x) = 0, \end{equation} where $ \kappa^2$ is the parameter with dimensions of area that appears in Einstein's equation.  $k^{\mu} k^{\nu} T_{\mu\nu} $ is the energy that is measured along the infinitely accelerated Unruh trajectory whose inflection point is the point $x$ and $k^{\mu} (x)$ is the limiting (past) tangent vector to that trajectory.  Since $k^{\mu}$ can be any null vector in the space-time, this gives us Einstein's equations except that it doesn't capture the c.c..  The meaning of this is the title of this section: the c.c. does not contribute to hydrodynamics.  

Instead, since the beginning of the century, Fischler and the author have championed an interpretation of the c.c. as a parameter determining the relation between the large proper time and large area limits of diamonds and, as a consequence of the principle of {\it asymptotic darkness}(AD)\cite{tbdg}\footnote{AD: The high energy behavior of scattering amplitudes (or their AdS equivalent) at impact parameters growing like the Schwarzschild radius of the center of mass energy, is dominated by black hole production (and decay in the Minkowski case). In the dS case, the c.c. determines the largest allowed black hole mass.}, the high energy behavior of the theory.  It cannot be renormalized by local physics, particularly low energy local physics.

Left to the imaginative effective field theorist reader of this essay are the implications of this point of view for much of the work on the c.c. in effective field theory.  Instead of delving into that, we note that the principle $S = A/(4 A_P)$ with $S$ interpreted as the logarithm of the dimension of the Hilbert space associated with a causal diamond\footnote{This interpretation of the entropy follows from Jacobson's use of an infinite temperature accelerated trajectory in his derivation.} can resolve many puzzles in the effective field theory treatment of black holes and inflation:

\begin{itemize} 

\item It gives us a way to make non-singular statements about space-times with space-like singularities.  For example, the Big Bang is just the statement that the Hilbert space available to a detector along an FRW geodesic is small at the origin of proper time (the Hamiltonian is time dependent so there's nothing crazy about time beginning at a finite point in the past).  Classical gravitational field theory is hydrodynamics, and only relevant for systems of large entropy.  Similarly, the shrinking of area of diamonds as one considers later proper time intervals after crossing the horizon of a black hole is just the statement that more and more of the degrees of freedom with which a detector could have interacted if the black hole didn't form, have come into equilibrium with the horizon.  The singularity is reached when the Hilbert space available to the detector is too small for hydrodynamics to be a good description of quantum mechanics.  We've already seen how the Covariant Entropy Principle can provide a sensible interpretation of CDL transitions from dS to Big Crunch cosmologies when the effective field theory is above the Great Divide.  In that case also, the singular region is one where all degrees of freedom in a Big Crunch diamond are equilibrated with the much larger Hilbert space of the dS progenitor, so that no effective field theory description in the Big Crunch region is valid.

\item For black holes in dS or Minkowski space, the process of a small low entropy system falling into, and becoming equilibrated with, a large black hole, creates a huge amount of entropy.  One interprets this as the statement that the Hilbert space of the combined system, even before infall, is much larger than the tensor product of the Hilbert spaces of the individual systems, but that degrees of freedom that mediate the interaction between the two systems are initially frozen into states that do not allow interaction\footnote{Think of systems whose variables are arranged into matrices with a single trace Hamiltonian.  The frozen degrees of freedom are off diagonal matrices.}.  The "almost flat space" experience of a detector right after falling through the horizon occurs during the long period of proper time that it takes to turn on these frozen variables and equilibrate the whole system.  This observation eliminates the so called firewall paradox\cite{amps}.

\item Empty dS space is a system with a finite Hilbert space.  Localized objects in that Hilbert space are constrained, low entropy, states of that Hilbert space.  There is no conventional notion of EI, with an ever expanding set of independent degrees of freedom in such a model.  The overwhelming majority of the states in the system cannot be associated with field theoretic degrees of freedom.

\item Models of inflation are not dS space.   They are models with a time dependent Hamiltonian, that initially couples together only the operators in a small finite dimensional Hilbert space, and then allows them to couple to a large set of degrees of freedom corresponding to causal diamonds in an FRW universe with no accelerated expansion for a long period of time.  In the penultimate section of this paper we'll review the Popular Science description of how this way of thinking about inflation leads to a post inflationary universe that is a dilute gas of black holes, with horizon size equal to the inflationary Hubble scale.  We'll see that such a model can account for much of the broad brush phenomenology of early universe cosmology, and possibly even predict that primordial black holes are part or all of the dark matter.

\end{itemize}

The overall message is that the boundaries of validity of effective field theory are reached when the localized entropy in a causal diamond in $d$ space-time dimensions, with size much smaller than the radius of curvature, exceeds $S_{CEP}^{(d - 1)/d}$.  $S_{CEP}$ is of course the entropy attributed to the diamond by Jacobson's principle, which is equal to Bousso's covariant entropy bound\cite{fsb}.  This bound on the regime of validity of quantum effective field theory is a covariant version of the cutoff of\cite{ckn}.  The underlying quantum gravity dynamics implied by this principal is radically different than effective quantum field theory with any standard cutoff.  The way in which effective field theory emerges from it has only begun to be understood\cite{tbwfsmatrix2}.  

\section{A Theory of Inflation Based on the CEP}

Let's begin by explaining why localized objects in dS space should be thought of as constrained states of a high entropy system called empty dS space.  By extrapolation to vanishing c.c., the same must be true of Minkowski space.  In Appendix B we'll explain how the same conclusion about Minkowski space can be drawn from the AdS/CFT correspondence.  The metric of a black hole in dS space, in Planck units, is
\begin{equation} ds^2 = - f(r) dt^2 +\frac{dr^2}{f(r)} + r^2 d\Omega^2 . \end{equation} 
\begin{equation}- r f(r) R^2 = (r - R_+)(r - R_-)(r + R_+ + R_-) . \end{equation}  $R$ is the dS radius.  The  parameter $R_+R_- (R_+ + R_-)$ is $2MR^2 $, the black hole mass in the limit $R \rightarrow \infty$ with $r\ll R$.  When $R_- \ll R_+$ we have $R_- \approx 2M$ , $R_+ = R - MR $.   Thus, the area of the cosmological horizon decreases by $ - 2\pi R M$, which is what we would expect for the probability of having an energy $M$ at temperature $(2\pi R)^{-1}$.  This is the Gibbons-Hawking temperature computed from quantum field theory in dS space, or looking at the Euclidean periodicity of the solution.

Our universe seems to be ending up as dS space, so all the things around us must have been created by some low entropy constraint on the initial conditions.  When $R$ is large, it is a {\it very} low entropy.   Boltzmann and Penrose, independently, worried about why the universe started off in such an "unnatural" state.   Fischler and the author call the explanation of this fine tuning problem the "localthropic principle".  It seems fairly evident that no complex organized system could ever evolve (except by improbable random fluctuation) in a completely randomized system with average energy $\sim 1/R \sim 10^{-42}$ GeV.  We need local objects in the universe in order to have anything interesting happen.   The question is, what is the most probable way to find localized objects, and how will they evolve?

In order to go further we have to say at least something about the HST formalism which has been proposed as the replacement for quantum field theory\cite{hst}.  We'll say as little as possible, so as not to get bogged down in the details.  The basic idea is that a nested set of causal diamonds, each having slightly larger area than the previous one, is equivalent to a nested set of intervals along some time-like trajectory.  According to the CEP this suggests that one should assign a quantum system to each time-like trajectory in a space-time.  In general, the causal diamonds along two different trajectories have an overlap.  CEP implies the maximal diamond in the overlap is a tensor factor in each trajectory's Hilbert space.  Dynamics along each trajectory prescribes a density matrix for this Hilbert space.  The {\it Quantum Principle of Relativity} says these overlap density matrices should be unitarily equivalent.  This is an infinite set of constraints for every possible overlap of causal diamonds along different trajectories.  

Causality implies that the time evolution operator along an interval must be the product of a unitary operator in the Hilbert space of the diamond of that interval times one that acts in the tensor complement of that space in the full Hilbert space. Correspondingly the Hamiltonian must be time dependent $H(t) = H_{in} (t) + H_{out} (t)$ where each piece is composed of operators acting in one factor of the tensor product.  This should not be a surprise: time slicings that stay within a causal diamond are, even in Minkowski space, never the same as those generated by an isometry of the global spacetime (see Figure 3 - which for future reference is drawn, for a general flat FRW model, in the Minkowski space to which the FRW model is conformal).    Causal time slices interpolate between different diamonds in Figure 3.  They are analogous to the Milne coordinates on the backward light cone of a point in Minkowski space.

\begin{figure}
\begin{center}
\includegraphics[scale=0.5]{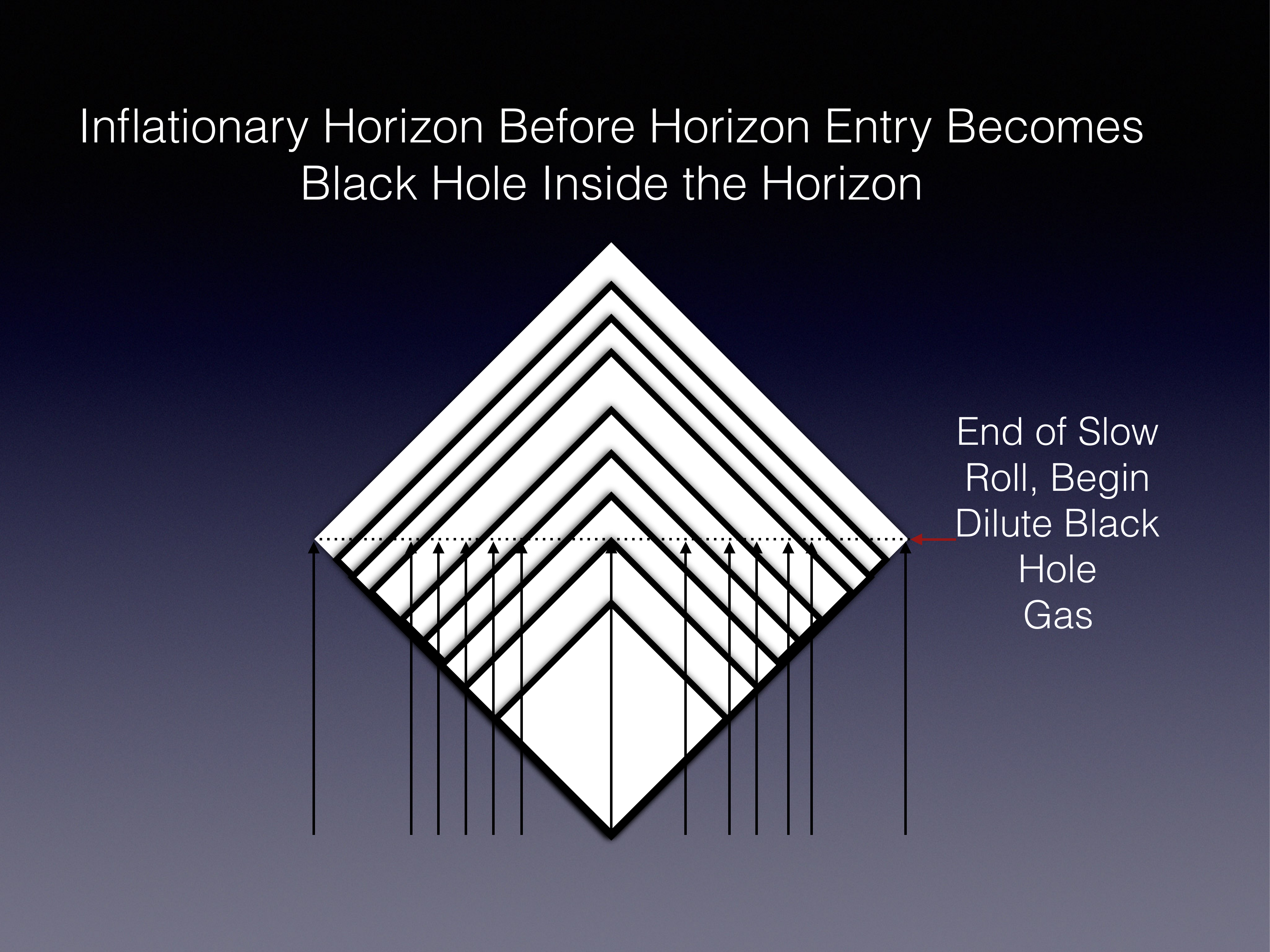}
\caption{HST Cosmology in conformal coordinates.  Black holes entering the horizon of one geodesic are inflationary horizon volumes along the geodesic intersecting horizon crossing}
\label{l}
\end{center}
\end{figure}

The philosophy of this approach, following the CEP, is that a classical space-time gives a hydrodynamic description of this whole family of quantum systems, telling us about the entropy in each causal diamond along a "sufficiently rich" set of trajectories.  Sufficiently rich means that the areas and intersections of all their causal diamonds (at discrete Planck time steps) tell us enough information to determine the space-time geometry with an accuracy on scales larger than the Planck scale. The task is then to find quantum systems that match this hydrodynamics, with the Quantum Principle of Relativity telling us how to knit the different  systems together consistently.  In particular, over each time interval, the $H_{out} (t)$ of one trajectory is determined by the $H_{in} (t)$ for trajectories whose causal diamonds are spacelike separated from that of the original trajectory.  

For a model of a homogeneous isotropic flat cosmology, we choose trajectories that are timelike geodesics that begin on a Planck scale regular lattice on the Big Bang hypersurface.  We note that the choice of a homogeneous flat cosmology is the specification of a model, rather than an initial condition.  We'll see that the initial conditions can be as general as we like, in each of the quantum systems, subject to the constraints of the quantum relativity principle.  Thus, in this way of looking at the quantum mechanics of the universe, the traditional flatness, homogeneity and isotropy problems are not problems of the initial conditions.  One could ask whether there are other, non-homogeneous, models, which could satisfy the consistency conditions, but even if there were, there would be no puzzle about fine tuning.  The traditional role of theoretical physics has been to find a mathematical model which fits the data of the universe we live in.  It was only during the second superstring revolution, as a consequence of the {\it incorrect} idea that all versions of string theory were somehow different states in a single quantum model, that physicists began to imagine that we could find the theory of the real world as a unique mathematically consistent model of quantum gravity, or the most probable state of a single model, which supports "intelligent life".   If there are different mathematically consistent models of quantum gravity, it is our task to find the one that fits the data we see and makes correct predictions about things that will be seen in the future.  It is not our task to explain why we are living in a world described by a particular model.

In the HST formalism, homogeneity is enforced by insisting that the time dependent Hamiltonians along each FRW geodesic are identical, up to a unitary transformation, and that the initial conditions are such that the density matrix consistency conditions can be satisfied. Isotropy is more complex and we won't need the technical details of it for most of the discussion. The only place it comes into our discussion of cosmology is in explaining the approximate $SO(1,4)$ invariance of the fluctuation spectrum.  Readers will have to consult\cite{revised} for that argument.

Now lets look at a particular geodesic, which we'll simply call {\it our geodesic}, at some instant of proper time when it can detect a local object entering its past horizon.  This is illustrated in the Figure 3.  A localized object is a quantum system which has not interacted previously with the degrees of freedom in the diamond, and will not interact with most of the horizon degrees of freedom until it exits the horizon.  Thus there must be constraints, setting to zero those variables that mediate the interaction.  The simplest and probably the only way to achieve this, consistent with the relativity principle, is to assume that along the geodesic that goes through the event of horizon crossing, that localized system has been evolving independently.  That is to say, along that geodesic, until the event of horizon crossing, that system has had a causal diamond of fixed area.   If the time from the Big Bang to horizon crossing is much longer than the Hubble time associated with that area diamond, then it must look like a typical state in empty dS space, with that dS horizon area.   In other words, up until that time, along that trajectory, the universe was inflating.  The homogeneity of our model then implies that on the FRW time slice corresponding to the time of horizon crossing, the trajectory that sees this local object must also have been inflating.  Thus, the HST cosmology assumes homogeneity and isotropy and derives inflation from the requirement that there are localized objects in the universe\footnote{HST also allows for the construction of a model in which there are never any localized objects.  The hydrodynamics of this model is a flat FRW model with scale factor $a(t) = \sinh^{1/3} (3t/R)$ .  It's an exact solution to the Friedmann equation with a mixture of $p = \pm \rho$ matter.}.  In this model, inflation lasts exactly half the conformal time $\eta_0$ between the Big Bang and dS singularities of the conformal coordinates.   In the HST model, neither of these singularities is singular (the dS singularity is of course just a coordinate singularity).  They are just points of minimal and maximal area causal diamond along a time-like geodesic starting at the Big Bang and going on to infinity.   The number of e-folds of inflation of course depends on the scale factor describing the entire hydrodynamic history of the universe.

We've seen that a localized excitation coming into the horizon looks like an inflationary horizon volume along the timelike geodesic that hits that event.   From the point of view of the original trajectory, on the same causal time slice but much later in FRW time it is a finite quantum system with entropy of order $H_I^{-2}$, which is decoupled from all of the rest of the degrees of freedom in the full Hilbert space.   This requires of order $H_I^{-1} R$ constraints on the final dS causal diamond, where $R$ is the ultimate dS radius.  This means the excitation carries an energy $H_I^{-1}$ .   That is, this system is a system with the energy and entropy of a black hole of mass $\sim H_I^{-1}$.  If we now look at all those localized excitations that come into the horizon over the lifetime of the universe, they form, on the FRW time slice at half the conformal time of the universe, a dilute gas of black holes.  

Implicit in this picture is a slow roll metric, which interpolates between the inflationary era and the dilute black hole gas era.   This is necessary because we have assumed that the horizon expanded between these two eras, so that the detector located on our trajectory at some conformal time $> \eta_0 /2$\footnote{The time in the system of causal coordinates for our trajectory's diamond is chosen to coincide with the FRW time on the FRW slice that intersects the trajectory.  That is, our geodesic is one of those defining the standard FRW coordinate system. From the point of view of our trajectory, other simultaneous points in causal coordinates are in the FRW past.} can receive signals from the inflationary horizon located at a point spatially separated from our trajectory in FRW coordinates.  These points are, of course, in the FRW past.  In HST, this expansion of the horizon is encoded in the way that variables are moved from $H_{out} (t)$ to $H_{in} (t)$ as time increases.  Inflation corresponds to a fixed Hilbert space for $H_{in}$ over some period of time, while slow roll means that we are slowly increasing the size of the $H_{in}$ Hilbert space and decreasing that of $H_{out}$ in order to keep the total dimension of the tensor product equal to the exponential of one quarter of the area of the cosmological event horizon of our asymptotically dS universe.  Note that any choice of slow roll is consistent with causality and unitarity.   We have not yet been successful in incorporating into HST the quantum principle of relativity for trajectories that are in relative motion.  While it's conceivable that this can put restrictions on the slow roll era it's hard to see how it might restrict the slowness of the slow roll.  Thus, it seems unlikely that this formalism will obey the restrictions on inflation that come from assuming a conventional field theoretic formalism with an inflaton potential constrained by {\it e.g.} Weak Gravity\cite{wgc} bounds.

Our conjectured entropy bound on the validity of effective field theory, $$S_{QUEFT} < A^{3/4} ,$$ implies that effective field theory is not valid during the inflationary and slow roll eras.  We have constructed explicit quantum mechanical models\cite{revised} which behave in the manner outlined above, but the equation of state relating the pressure and energy density of the slow roll era does not correspond to any sort of localized excitations. 
Note that despite this, if we {\it define} the pressure and energy density from the slow roll metric using the Friedmann equation, then as long as $p + \rho \geq 0$ we can invent a fictional "inflaton field" via the equations
\begin{equation} p = \frac{\dot{\phi}^2}{2} - V(\phi ) , \end{equation}
\begin{equation} \rho = \frac{\dot{\phi}^2}{2} + V(\phi ) . \end{equation} and fit $V(\phi)$ to the chosen scale factor $a(t)$ .   Since $\phi$ is a purely hydrodynamic construction there is no reason for it to be the proper quantum variable to use in this high entropy density situation.  In the HST formalism it is not.  

In our work on this model, we've made the assumption that the new degrees of freedom that come into the horizon during slow roll, after the exit from a period of inflation with fixed Hilbert space, do not equilibrate with the original black hole system.  We know that a model in which the degrees of freedom of the entire Hilbert space are constantly kept in equilibrium, saturating the covariant entropy bound at each moment, actually has the equation of state $p = \rho$ and is not a slow roll.
So it seems reasonable to assume that the slow roll does not satisfy the adiabatic condition  for the Hubble parameter $h(t)$, that is, we should require
\begin{equation} \dot{h} > c \frac{h^2}{{\rm ln}\ h} ,  \end{equation} with $c = o(1)$.  The value of $c$ would depend on the numerical coefficients in the scrambling rate in the underlying quantum mechanics.  We do not yet know if there are any {\it a priori} bounds on it. The meaning of this equation is that the black holes that come into the horizon at some late FRW time on our trajectory, do not, in the dynamics along their own trajectory, equilibrate with the growing horizon during the slow roll era.  If they did we would not obtain a dilute gas of black holes on the FRW slice at $\eta_0 / 2$ but instead find that the horizon at that time was filled by a single black hole.  Thus, the choice of slow roll metric satisfying the inequality $\epsilon > c {\rm ln}| h$ is the mechanism which gives the "phase transition" between the "Dense Black Hole Fluid" that saturates the covariant entropy bound at all time, and the dilute black hole gas.

On the FRW time slice at the end of the slow roll era, the universe is filled with a dilute gas of black holes.  These black holes have the same size, on average, but they are finite quantum systems in equilibrium and thus have approximately Gaussian\footnote{This is just the central limit theorem of statistical mechanics.} fluctuations of all their macroscopic properties, whose size is of order $S^{-1/2}$ where $S$ is the entropy.   Although the HST rules for an FRW universe say that the Hamiltonian is the same along all FRW trajectories the initial states are constrained only by the density matrix consistency conditions for overlaps.  Furthermore, on the FRW time slice corresponding to conformal time $\eta_0 / 2$ the black holes all came into the horizon of any particular geodesic at different proper times along the geodesics that go through the horizon crossing events.  Thus, a late time detector will see the fluctuations as fluctuations of the energy density on the FRW time slice.  As usual, in the gravitational field equations that describe the hydrodynamics of the system, one can switch to global comoving time slices on which the energy density and pressure are constant, and the fluctuations are fluctuations of the scalar curvature $h$.   

On comoving slices, the invariant measure of scalar fluctuations is \begin{equation} \zeta = h \delta \tau = \frac{h \delta h}{\dot {h}} = \epsilon^{-1}  \frac{\delta h}{h}, \end{equation} where $\epsilon$ is the usual slow roll parameter.  The HST prediction is 
 \begin{equation} \frac{\delta h}{h} \sim n^{-1}, \end{equation} where $n$ is the average black hole radius, which is $H_I^{-1}$ in Planck units. $H_I$ is the inflationary Hubble parameter. There are two differences between this formula and the formula predicted by single field slow roll inflation. First of all, there is no absolute normalization for the fluctuations, since everything we've said is valid for a large class of choices for the underlying time dependent HST Hamiltonian, whereas single field inflation has predictions that depend only on the Einstein action and the choice of slow roll metric\footnote{The inflaton field is just as much a fiction in single field inflation as it is in the HST formalism.} .  Secondly, in single field inflation one predicts 
 \begin{equation} \frac{\delta h}{h} \sim \sqrt{\epsilon} h^{-1} (t) . \end{equation}  The extra time dependence, relative to the HST prediction, has to do with the normalization of the field $\zeta$, viewed as a quantum fluctuation of the gravitational field in comoving coordinates, and the fact that those fluctuations are computed in the time dependent slow roll metric.   We stress the fact that those predictions are calculated using quantum field theory in a regime where the entropy in the inflationary causal diamond is assumed to saturate the covariant entropy bound, in violation of our rule that one should believe field theory only in regimes where the localized entropy satisfies
 \begin{equation} S_{loc} \leq S_{CEP}^{3/4} . \end{equation}
 
 Despite the difference in the formulae, there is a simple criterion that makes these two sets of predictions indistinguishable in current data.  This is the assumption that the spacetime variations of the two point fluctuations are approximately $SO(1,4)$ invariant.  The justification of this assumption within the HST formulation requires more technical detail than would be appropriate in this brief review but can be found in\cite{revised} .  The essential point though is that we currently have no other observational constraints on the slow roll metric, so that in either formalism we are simply fitting that metric to the data.  So far, the only theoretical constraint we have on this metric is the inequality
 \begin{equation} \epsilon > c/({\rm ln}\ h(t)) , \end{equation} saying that most of the extra states appearing inside the horizon during the slow roll era expansion of the Hilbert space, become part of the cosmological horizon.  This is equivalent to saying that we want to minimize the number of constraints on the initial state compatible with having enough localized entropy in the universe to have a radiation dominated era.  The constant $c$ depends on the detailed choice of microscopic Hamiltonian in HST.
 
 The predictions of the two classes of models for non-Gaussian fluctuations and the tensor scalar ratio are quite different, but if $\epsilon$ is small both frameworks predict that these effects are small.  Quantitative estimates of how small depend, in HST, on details over which we presently have no control.  HST gives a qualitatively new account of the reheating era, which occurs through black hole decay. Contrary to popular opinion, once one takes into account the variation of the black hole mass upon decay, which is significant for the tiny black holes of the HST model, these black hole decays can lead to baryogenesis, if one assumes a reasonable amount of CP violation in the decay amplitudes.  The holes have horizon size $\sim 10^6$ Planck lengths if one fits the model to inflationary fluctations, using $\epsilon \sim 0.1$.   
 
 One predicts a reheat temperature $\sim 10^{10}$ GeV if the black hole number density on the $\eta_0 / 2$ surface is just below that which would cause them to coalesce into a $p = \rho$ state, but instead allows them to behave like a $p = 0$ gas.  The $p = \rho$ state saturates the covariant entropy bound, and so corresponds to a state with few or any constraints, while the gas implies more constraints on the boundary degrees of freedom and so is {\it a priori} a less probable initial condition.  We don't know how to put a precise value to the inequality "just below the transition between $p = \rho$ and $p = 0$" , but it is presumably a density of order $\frac{d_0}{n^3}$ with $d_0$ of order $1$.  The immediate post-inflationary universe is thus matter dominated, with an initial black hole number density $\frac{d_0}{n^3}$, at conformal time $\eta_0 / 2$.
 The density fluctuations then grow to order $1$ when the scale factor has grown to
 \begin{equation} \frac{a(\eta )}{a(\eta_0 / 2)} = \bar{\epsilon} n . \end{equation}  In this equation $\bar{\epsilon}$ is an average value of the slow roll parameter during the slow roll era.
 At this FRW time, the energy density is 
 \begin{equation} \rho (\eta ) = \frac{d_0}{( \bar{\epsilon}n^2)^3} = d_0 \bar{\epsilon}^2 10^{- 25} , \end{equation} in Planck units.  This is $10^9$ times larger than the energy density of reheating by black hole evaporation, so there is plenty of time for some of the black holes to coalesce.   The determination of the spectrum of black hole masses  at reheating is a crucial parameter for understanding the phenomenology of the HST model.  Black holes whose Hawking temperature is lower than the reheat temperature continue to grow during the radiation dominated era.   These are black holes that are about $10^3$ times larger than the initial post-inflationary holes.  One must then calculate the distribution of masses at the end of the radiation dominated era.  One could find black holes that would evaporate during the matter dominated era, which are likely to be ruled out by observational constraints.  Alternatively, one might find a distribution of cosmologically stable holes, most likely within a few orders of magnitude of the cosmological stability bound, around $10^{22}$ Planck masses, since that is already many orders of magnitude larger than $10^9$.  We note that, due to recent progress in understanding small scale gravitational lensing, there is a window of a few orders of magnitude above the cosmological stability bound, where primordial black holes could constitute all of the dark matter. 
 
 To summarize:  the barest bones of the ideas behind the HST models of quantum gravity lead to an early universe cosmology radically different from that based on field theoretic inflation, but equally consistent with current data.  The model is completely finite, even at the Big Bang "singularity", and provides an extremely economical,almost purely gravitational, account of inflation, reheating, baryogenesis, and possibly primordial black hole dark matter.  The biggest uncertainty in the model, as in conventional inflation, is the form of the slow roll metric that interpolates between the inflationary era and a primordial black hole dominated era.  There appear to be no constraints on that metric from unitarity or locality/causality, though there might be some coming from the (so far unimplemented) HST consistency conditions relating timelike trajectories in relative motion.   All HST models of cosmology predict that a universe with any localized excitations has constrained low entropy initial conditions\footnote{Indeed, this conclusion follows from the assumption that we are approaching a long lived dS era, and the black hole entropy formula in dS space.}.  We get some constraints on the slow roll metric by requiring the most probable initial conditions that lead to a radiation dominated era.   The alternative is a universe with either no local excitations or a non-homogeneous collection of large black holes.
 
 By far the strongest constraint on the slow roll era is simply fitting the CMB spectrum, which means that in the HST framework, this spectrum is not really predicted.  This is really the {\it same} situation that contemporary field theoretic models of these fluctuations are in.  In the field theoretic framework, one fits the inflaton potential to the data and hopes at some future date to calculate that potential from a UV complete theory.   HST is a UV complete theory, but so far does not provide either theoretical or environmental selection criteria for predicting the detailed form of the scalar CMB two point function.  We will have to wait for positive signals for tensor fluctuations and/or non-Gaussianity to differentiate between the two models.
 
 Even if HST is completely wrong, there is an important lesson to be learned from this.   Jacobson's 1995 article identifies the general theory of relativity as hydrodynamics of the area law for causal diamonds.   There is a general theorem, the {\it fluctuation dissipation theorem}, which implies that hydrodynamic equations are really stochastic equations.  Recent derivations\cite{hongrangamanitblucas} of hydrodynamic equations directly from the Heisenberg equations of quantum systems, show that hydrodynamics arises as a set of stochastic differential equations.  No inflation theorist denies that the inflationary era is a high entropy situation.  Thus, one should expect statistical fluctuations of the gravitational field of order $H/M_P$ the inverse square root of the entropy.  Impose the symmetry constraint of de Sitter invariance on the power spectrum of the curvature fluctuations, but recognize that an end to inflation is necessary to make those fluctuations visible as spatial fluctuations on the sky.  The gauge invariant measure of scalar fluctuations is related to curvature fluctuations in comoving gauge by a factor of $1/\epsilon$, and so is not exactly dS invariant.  This bare bones framework is enough to explain all the data we have so far on the inflationary era.  We can fit the slow roll metric to the deviations of the CMB power spectrum from scale invariance.  We can, as in field theory models of inflation, insert dependence on the slow roll metric into the curvature fluctuations themselves.  That will change the formulae, but not our ability to fit the Gaussian power spectrum.
 The smallness of tensor fluctuations and non-Gaussianity is explained by the same factor of $1/\epsilon$, plus approximate dS invariance and Maldacena's squeezed limit theorem.   Note that that theorem is symmetry based and so is valid beyond the quantum field theory context that was used to derive it.   Until we measure tensor fluctuations or non-Gaussianity, we will learn nothing more from observation about the underlying quantum system that is responsible for what we have observed so far.
 The success of QFT in fitting extant data means only that QFT obeys these general rules.

\section{Conclusions: Executive Summary}

The most important lessons to be learned from the arguments of this paper are
\begin{itemize}

\item As field theorists we are used to thinking of a low energy effective Lagrangian as a description of the the low energy sector of a particular quantum mechanical model.  Every low energy classical solution of it corresponds to a state in the same Hilbert space, modulo the infrared decoupling that sometimes occurs in the limit of infinite volume.  In models of quantum gravity this is manifestly incorrect.  Different low energy solutions can correspond to different models with radically different {\it high} energy behavior.  One low energy solution might correspond to a well defined QG model, while the other is in the Swampland.  The evidence for this contention comes from multiple independent sources: the fact that Hamiltonians in generally covariant theories are defined on the asymptotic boundaries of space-time, black hole physics, CDL tunneling solutions, perturbative string theory, Matrix Theory, and AdS/CFT.  A corollary of this conclusion is that it is not sensible to compute quantum corrections to the effective action that matches the low energy observables of one theory of QG, and then use that calculation as evidence for another model with very different asymptotic behavior in space-time.  

\item There is no evidence in either QFT in dS space, or from CDL tunneling solutions, for the radical theory known as Eternal Inflation.  In QFT one can make a rigorous statement.   The only non-singular dS invariant state for any QFT is the BD vacuum defined by the Euclidean functional integral on the sphere.  Its predictions for potentials with multiple minima are thermal.  They are finite and unambiguous.  States corresponding to tests of the hypothesis of EI by making measurements of independent fluctuations on late time slices in any global slicing of dS space are slice dependent.  In the limit that the global time is taken to infinity none of these states lie in the BD Hilbert space, and they would all be singular on the cosmological horizon of a single causal patch.  These conclusions are consistent with the fact that the space of collective coordinates of field theory instantons on Euclidean dS space is compact.
CDL transitions provide further evidence against EI, because they provide evidence that the Hilbert space of states describing dS space is finite dimensional.  

\item More precisely, CDL transitions divide the space of effective local actions into two classes.
Given a dS minimum, one defines the classes by adding a negative constant to the action to bring that minimum down to zero and asks whether that Minkowski solution has a positive energy theorem.  If it does, the CDL transitions are consistent with the contention that the dS minimum describes a quantum theory in a finite dimensional Hilbert space with most of the states having energy below the dS temperature.  The backward and forward transitions are related by the principle of detailed balance, with the entropies defined by the areas of maximal causal diamonds in the two Lorentzian continuations of the instanton.  Following the actual classical evolution of the CDL Big Crunch we find that it leads to a high entropy situation in which the scalar fields explore their whole configuration space and excite non-homogeneous field theory modes until the field theory state violates the entropy bound.  It's thus reasonable to assume that the system will settle down into its maximal entropy state, which is the original dS minimum.  All of these transitions are consistent with a model of dS space with a finite number of quantum states.

\item Below the Great Divide, CDL bubbles do not swallow generic excited states of the the "unstable" Minkowski solution, but are instead swallowed by them.  It's plausible that all of these effective field theory models are in the Swampland. 

\item The only model of CDL tunneling that indicates a true decay of dS space are transitions into zero c.c. FRW models.  There does not yet exist a quantum theory that reproduces the physics of these models.

\end{itemize}

The standard paradigm of effective field theory is that an effective Lagrangian is obtained by integrating out high energy short distance degrees of freedom.  The conclusions above suggest that this is incorrect, and black hole physics tells us that QG has a short distance cutoff at the Planck scale.  In space-times with non-negative c.c. and causal diamonds of finite area, once the field theoretic entropy exceeds $A^{\frac{d -1}{d}}$ , the maximal entropy states are black holes.  Large black hole decays are sensitive to the detailed low energy spectrum of the model.   This long distance/high energy connection implies that we must re-evaluate the way in which field theory emerges from models of QG.  There are two more observations about horizon entropy, which are guides to what the correct theory looks like:

\begin{itemize}

\item  According to the Schwarzschild dS black hole entropy formula, inserting a small localized object of energy $E$ into empty dS space {\it reduces} the entropy by an amount $2\pi R E$.  This is a purely classical, up to the basic covariant entropy postulate, derivation of the Gibbons-Hawking temperature of dS space, originally derived using quantum field theory.  This strongly suggests that empty dS space is a thermalized finite dimensional system with all states localized on the cosmological horizon and having an energy $< 1/R$.  Localized objects are constrained states of this ensemble and their energy $E$ parametrizes the the number of constraints.  Energy is only defined up to corrections of order $1/R$.  These peculiar sounding results sound more sensible when we realize that we are talking about energy in a static coordinate system. In such a coordinate system, nothing ever falls through the horizon.  Instead objects near the horizon have redshifted energy that goes to zero the closer they are to the horizon.  Furthermore, a localized object sitting exactly on the timelike geodesic whose causal patch is covered by the static coordinates, will be struck by Gibbons-Hawking gravitons in a time of order $1$ and migrate to the horizon because of the "Hubble flow" in a time proportional to $R$\footnote{It's important to note that this is a description of what happens from the point of view of an idealized detector traveling on a geodesic. Along the object's own trajectory through space-time the effect of the Gibbons-Hawking gravitons is negligible for much longer times.}.

\item If one drops a small object of mass $m$ onto a large $4$ dimensional Schwarzschild black hole there is a huge increase in entropy of the eventual thermal equilibrium state, by an amount $8\pi R_S m$.   This indicates that the system just before the object passes through the horizon lives in a much larger Hilbert space than can be accounted for by the pre-existing black hole and the individual system and that the state before the object crosses the horizon is a constrained state of that larger Hilbert space, with a constrained subspace of entropy $8\pi R_S m$ .   The energies of the states involved must be $< 1/R_S$ in order not to change the classical energy of the system.  The time scale of equilibration is thus of order $R_S$.  The QFT model of black hole entropy in terms of near horizon states of short wavelength field oscillations, with 't Hooft's "brick wall" cutoff, cannot account for this change in entropy.   Even in terms of energetics, the brick wall model fails to account for the entropy as seen by a geodesic observer.  In QFT the Hamiltonian for a geodesic coordinate system like Novikov coordinates has no low energy states besides the vacuum.  

\item A class of models that {\it do} account simply for this increase in entropy are matrix models with single trace Hamiltonians
\begin{equation} (R_S + 2m)^{-1} {\rm tr }\ P(\frac{M}{R_S + 2m}), \end{equation} with polynomial $P$
.  Each matrix element of the $(R_S + 2m)\times (R_S + 2m)$ matrix  $M$ is an operator in a small finite dimensional Hilbert space.  The constrained, pre-horizon crossing state has vanishing matrix elements for the two off diagonal $R_S \times 2m$ blocks.  These models are all {\it fast scramblers} and so account for the Hayden-Preskill\cite{hpss} observation about scrambling times for black holes.  They resolve the firewall paradox posed by the single low energy quantum state postulated for geodesic Hamiltonians in QFT.

\end{itemize}

These two observations can be generalized to higher dimensions, and demonstrate that QFT has infrared as well as UV deficiencies, when viewed as an effective low energy approximation to a consistent model of QG.
Together with the CEP, these two arguments lead to an approach to local theories of QG which is not based on QFT.   The Hamiltonians are single traces of matrices. In order to incorporate the right entropy counting and dynamical properties of higher dimensional horizons, the matrices are constructed as bilinear contractions of $d-2$ dimensional tensors, with a factor of $1/N$ relative to the large $N$ scalings of \cite{guraurivasseau}.  That factor ensures that horizon degrees of freedom have a natural time scale $R_S$ when $N^{d-2}$ is set proportional to the horizon area.  
The individual tensor elements are operators in some fixed Hilbert space with a dimension $d_C$ independent of $N$.  The total entropy is then $N^{d-2} {\rm ln} d_C$ .   Thus, the log of the dimension of the "internal" Hilbert space contributes a multiplicative factor in the Planck scale, in a manner reminiscent of Kaluza-Klein theories.   The HST formalism makes more specific choices about the nature of the operator algebras involved, motivated by the ubiquity of SUSY in string theories, but very little of that structure has been used so far in the application of the formalism to cosmology.

The purpose of this note is not to extol the virtues of the HST formalism, or examine its glaring faults, but rather to point out to effective field theorists that the ideas of Eternal Inflation, the String Landscape, and the possibility of computing low energy corrections to the c.c. have no well founded theoretical basis.  They are based on the idea that the relation between QFT and a complete theory of QG is similar to that between QFT and a lattice Hamiltonian.  Classical black hole physics, CDL instanton computations, and manifold examples from both perturbative and non-pertubative string theory, show that the last of these ideas is definitely incorrect, and that the other two have not been validated by actual computations, but only by conjectures based on effective field theories derived from string theories.  In the case of EI, even the effective field theory description is incorrect. The key lesson that different solutions of the same gravitational effective Lagrangian, with different large time asymptotics, represent different models of QG, rather than different states in the same model, shows that arguments for the landscape and for tunneling EI are far from persuasive.   

Instead we've argued that the covariant version of the CKN\cite{ckn} bound on effective field theory, $S_{QUEFT\ \Diamond} < A_{\Diamond}^{\frac{d-1}{d}}$ and the BHtHJFSB\cite{bhthjfsb} Covariant Entropy Principle ${\it ln\ dim} {\cal H}_{\Diamond} = \frac{1}{4} A_{\Diamond}$\footnote{Both of these bounds apply only to regimes where the area is large compared to any microscopic scale.  In weakly coupled string theory they will have corrections that are model dependent.}  serve as our best guides to the regime of validity of Quantum Effective Field Theory.  Jacobson's Principle tells us that classical field equations give us a coarse grained hydrodynamic description of QG, even in regimes where QUEFT is not valid.  These principles can help us to understand singular solutions of GR, like the interior of black holes, the Big Bang, and the Big Crunch of CDL transitions for Lagrangians above the Great Divide, as representing low entropy subsystems in a model of QG.  The singularities in these classical geometries correspond to points in time at which those subsystems have too few degrees of freedom to be described by hydrodynamics.  For the Big Bang singularity this occurs simply because of the constraints of causality and the CEP, while for black holes and CDL crunches, it occurs because the degrees of freedom of the temporarily independent subsystem have come into equilibrium with a larger system, so that a hydrodynamic description of the subsystem is no longer valid.

HST was invoked in this article only to demonstrate that the observational phenomenology of "inflation" could be reproduced by a model with no quantum fields, that was compatible with the principles of unitarity and causality.  That model is incompatible with the idea of EI for an asymptotically dS universe, and does not seem to obey the WGC restrictions on the flatness of the inflaton potential.  It has testable predictions for tensor fluctuations and non-Gaussianity, that are different from those of any field theory based model.  The more important observation that the HST model highlights is that all of the existing data on "inflation" can be explained by a much less microscopic interpretation of the symmetry constrained statistical hydrodynamic Einstein equations, without resorting to an underlying quantum model.

There is no denying that there is a regime of phenomena for which effective quantum field theory is an excellent approximation.  It has been tested experimentally with high precision.  Those phenomena do not include inflation, eternal inflation, landscapes, tunneling transitions between "vacua", or black hole physics.  Instead, everything we know about those phenomena is explainable by what one might call the statistical hydrodynamic interpretation of general relativity, and a few symmetry principles.  We have been fooled by relying too heavily on the detailed formalism of QFT in regimes where it has not been tested.

We will conclude with a list of issues raised in this paper, which deserve further investigation

\begin{itemize}

\item More work is necessary to understand whether there are consistent models of quantum gravity whose dynamics is well approximated by the CDL instanton for decay of dS space into a zero c.c. FRW spacetime.  It would probably also be a good idea for someone to revisit the examples of tunneling which can be modeled using the AdS/CFT correspondence.  Making sure that the various excellent papers on this are consistent with each other and finding further evidence for the picture that emerges from that literature would be a very useful exercise.

\item The authors of\cite{ckn} argued that their bound on the range of validity of EFT was consistent with all known calculations of radiative corrections.  This investigation should be redone and extended to questions about the use of QFT in contexts where it has not been tested.  Important examples are the QFT theory of cosmological phase transitions and the evolution of cosmic defects.  

\item Within AdS/CFT it is important to understand how states with infinite numbers of soft gravitons emerge in the infinite radius limit.  It's also imperative to understand whether it is always necessary to have compact dimensions whose size is of order the AdS radius, and whether supersymmetry, broken only by relevant operators, is necessary to the existence of the dimension gap which indicates an Einstein gravity dual to a CFT.

\item Within the context of HST the most pressing problem is to find Hamiltonians which satisfy the consistency conditions for trajectories in relative motion.  This is what guarantees the Lorentz invariance of the S-matrix for these models.

\item The HST model of AdS space leads to a general prescription for constructing tensor network lattice models, which converge to a given CFT.  The connection between tensor networks, proper time evolution in AdS space, and CFT needs much more work.  Large $c$ models in $1+1$ dimensions, which are permutation orbifolds of soluble models are a nice class of models to think about in this connection.  Probably none of these have large radius duals, but perhaps some of them can be related explicitly to string theory on background AdS space.  
\end{itemize}
\section{Appendix A: Large Black Holes in AdS Space and Large Compact Dimensions}

The metric, in global coordinates outside of a large AdS black hole is approximately
\begin{equation} ds^2 = - f(r)dt^2 + \frac{dr^2}{f(r)} + r^2 d\Omega^2 . \end{equation}
\begin{equation} f(r) \approx r^2/R^2 - 2M/r . \end{equation}

The Schwarzschild radius is approximately given by $R_S = (2MR^2)^{1/3} $.  The temperature is $T^{-1} = (M^{-2/3}/3)  \pi (2R^2)^{\frac{2}{3}}$ Far outside this radius, the coefficients of $dt^2$ and $dr^2$ scale the same as $r$ is increased.  A shortcut to understanding the properties of quasi-normal modes\cite{horhub} of these black holes is to use the AdS/CFT correspondence.  The black hole is dual to a thermal state of the CFT on the boundary manifold $R \times S^2$ with radius of the sphere of order $R$.  This is the scaling one needs so that boundary energies and bulk masses are the same order of magnitude.  In a thermal state in a CFT, information propagates ballistically and the time of decay of a local perturbation is of order the inverse temperature.  This means that the perturbation decays exponentially, but does not spread over the entire sphere by the time it has decayed to the size of a thermal fluctuation.  Rescaling this relation to the position of a detector orbiting in a stable orbit a few Schwarzschild radii from the horizon we see that the information is scrambled at the fast scrambling rate in a local time of order $R$, but spreads over distances $\gg R$ on the horizon, only in a ballistic manner.  This is consistent with a tensor network reconstruction\cite{swingleetal} of the region near the horizon, in terms of a lattice field theory with lattice spacing of order $R$.  The fact that the scrambling occurs rapidly at each lattice point implies that each point is a large entropy system with ${\rm ln}\ S \propto {\rm ln} R$.  In all well studied examples, the large entropy is associated with a "fuzzy" compact manifold of radius $\propto R$, and dynamics that is non-local on that manifold.   Since the behavior of quasi-normal modes depends only on the geometry of $AdS$ space, what we are saying is that consistency of AdS/CFT with that geometry implies something like large compact dimensions, in the sense that each point in $AdS$ space has a large number of degrees of freedom associated with it.  The non-local nature of the interactions can be understood in terms of Hamiltonians invariant under volume preserving maps.  Under such maps a localized region can be mapped into a "thin amoeboid" of equal area, so there is no sense in which information initially associated with a localized degree of freedom has to travel a certain distance in order to be transferred to some other degree of freedom.  Scrambling is faster than ballistic.

\section{Appendix B: Minkowski Space from AdS/CFT}

The seminal papers on deriving flat space scattering amplitudes from CFT correlators are\cite{polchsuss}.  Most of the subsequent literature is concerned with the technical details of calculating limits of Witten diagrams.  A key concept in\cite{polchsuss} is the {\it arena}, a causal diamond of area much greater than the Planck or string scale but much smaller than the scale of the AdS radius, and the key argument was that by delicate tuning one could arrange for the boundary correlators finite numbers of certain local operators, sums over many primaries, to make, in the Witten diagram approximation, localized wave packets in the arena.  The tuning is necessary because generic local operator insertions on the boundary will miss any given choice of the arena.  

There was a deficiency in the argument of\cite{polchsuss}, which has never been remedied.  All well studied examples of the AdS/CFT correspondence have a compact internal space with $\geq 2$ dimensions and linear size of order the AdS radius.  Thus the flat space amplitudes live in higher dimensions and one must study high dimension Kaluza-Klein states in order to compute anything beyond a $4$ point function.  This is particularly clear for $AdS_3$ models, since there are no flat space quantum gravity models in $3$ Minkowski dimensions.  Any non-zero subenergy in a would be scattering amplitude determines a different asymptotic geometry.  We will ignore this complication, but someone should really clear it up.

We will make one extra assumption, namely that the quantum information in the arena is encoded in the state of a finite dimensional {\it code subspace} whose dimension is determined by the area of the arena's holographic screen.  Consider the geodesic of AdS space, which connects the past and future tips of the arena. The proper time along this geodesic is the time coordinate in a global coordinate system corresponding to a particular choice of the generator $H \equiv K_0 + P_0$ in the conformal group.   Let's conjugate this generator to obtain
\begin{equation} H(r) = e^{irL} (K_0 + P_0) e^{- irL} , \end{equation} where $L$ generates translations on the negatively curved spatial slices orthogonal to the timelike geodesic that goes between the tips of the arena.  A generic state with average eigenvalue of $H(r) = T R_{AdS}^2$ is represented geometrically by a black hole of radius $4\pi R^2 T/3$ centered at $r$.  If $r \gg R_S$ this black hole does not overlap the arena and so the local state of the arena must be approximately the Minkowski vacuum.  However, in the error correcting code representation of bulk data on the boundary\cite{almheiriharlow} the code subspace is always maximally entangled with a much larger system.  Therefore, the Minkowski vacuum, as approximated by a large radius AdS space, must be a maximal entropy density matrix on the code subspace.  Localized excitations in the arena must therefore be constrained states of lower than maximal entropy on the code subspace. Thus, at least qualitatively, the realization of Minkowski space dynamics as a limit of AdS dynamics agrees with its realization as a limit from dS space.  

The infinite entropy, zero energy factor in the Minkowski Hilbert space should be interpreted as the space of "soft gravitons" encountered in Feynman diagram approaches to quantum gravity in Minkowski space.   We know that, at least in four dimensions, the scattering operator has no matrix elements between Fock space states, because of IR divergences.  Weinberg's analysis\cite{weinbergir} shows us that the Fock space states of particles with momenta above a "detector cutoff" are entangled with an as yet poorly defined space of "soft gravitons".  While there are no IR divergences in higher dimensions, Fischler and the author have argued\cite{tbwfir} that the failure of Borel summability in string perturbation theory, and general unitarity arguments, suggest that even in higher dimensions, the S-matrix is not unitary in Fock space\footnote{To make this claim precise, one must have a definition of the space of states in which it is unitary. Matrix Theory\cite{bfss} and HST each provide us with such a definition, and I've suggested an heuristic, Lorentz covariant, picture of this space in\cite{cacb}.}.

\begin{center} 
{\bf Acknowledgments }  \end{center}

This work is partially supported by the U.S. Department of Energy under 
grant DE-SC0010008.  I would like to thank Patrick Draper for one of the figures, and extensive comments, and Willy Fischler and Michael Dine for comments on the draft. Special thanks to E. Rabinovici, for an incisive critique. As indicated in the text, this work was written in response to the stimulating program at KITP-UCSB, on Vacuum Energy and Electroweak Fine Tuning, which I attended from July 20 - August 4, 2019.  I'd like to thanks the organizers for inviting me and the staff at KITP for their usual efficiency and kindness.

\end{document}